\documentclass[aps,prl,reprint,superscriptaddress,footinbib]{revtex4-2}

\usepackage[utf8]{inputenc}
\usepackage{xcolor}
\usepackage{graphicx} 
\usepackage{epsfig}
\usepackage{epstopdf}
\usepackage{braket}
\usepackage[normalem]{ulem}
\usepackage[colorlinks]{hyperref}
\usepackage{bm}
\usepackage{siunitx}

\begin{document}

\title{A telecom-wavelength quantum repeater node based on a trapped-ion processor }

\author{V.~Krutyanskiy}
\affiliation{Institut f\"ur Experimentalphysik, Universit\"at Innsbruck, Technikerstr. 25, 6020 Innsbruck, Austria}

\affiliation{Institut f\"ur Quantenoptik und Quanteninformation, Osterreichische Akademie der Wissenschaften, Technikerstr. 21a, 6020 Innsbruck, Austria}

\author{M.~Canteri}
\affiliation{Institut f\"ur Quantenoptik und Quanteninformation, Osterreichische Akademie der Wissenschaften, Technikerstr. 21a, 6020 Innsbruck, Austria}
\affiliation{Institut f\"ur Experimentalphysik, Universit\"at Innsbruck, Technikerstr. 25, 6020 Innsbruck, Austria}

\author{M.~Meraner}
\affiliation{Institut f\"ur Quantenoptik und Quanteninformation, Osterreichische Akademie der Wissenschaften, Technikerstr. 21a, 6020 Innsbruck, Austria}
\affiliation{Institut f\"ur Experimentalphysik, Universit\"at Innsbruck, Technikerstr. 25, 6020 Innsbruck, Austria}

\author{J.~Bate}
\affiliation{Institut f\"ur Experimentalphysik, Universit\"at Innsbruck, Technikerstr. 25, 6020 Innsbruck, Austria}

\author{V.~Krcmarsky}
\affiliation{Institut f\"ur Quantenoptik und Quanteninformation, Osterreichische Akademie der Wissenschaften, Technikerstr. 21a, 6020 Innsbruck, Austria}
\affiliation{Institut f\"ur Experimentalphysik, Universit\"at Innsbruck, Technikerstr. 25, 6020 Innsbruck, Austria}

\author{J.~Schupp}
\affiliation{Institut f\"ur Quantenoptik und Quanteninformation, Osterreichische Akademie der Wissenschaften, Technikerstr. 21a, 6020 Innsbruck, Austria}
\affiliation{Institut f\"ur Experimentalphysik, Universit\"at Innsbruck, Technikerstr. 25, 6020 Innsbruck, Austria}

\author{N.~Sangouard}
\affiliation{Institut de Physique Th\'eorique, Universit\'e Paris-Saclay, CEA, CNRS, 91191 Gif-sur-Yvette, France}

\author{B.~P.~Lanyon}
\email[Correspondence should be sent to ]{ ben.lanyon@uibk.ac.at}
\affiliation{Institut f\"ur Experimentalphysik, Universit\"at Innsbruck, Technikerstr. 25, 6020 Innsbruck, Austria}

\affiliation{Institut f\"ur Quantenoptik und Quanteninformation, Osterreichische Akademie der Wissenschaften, Technikerstr. 21a, 6020 Innsbruck, Austria}

\date{\today}

\date{\today}

\begin{abstract}

A quantum repeater node is presented based on trapped ions that act as single photon emitters, quantum memories and an elementary quantum processor. The node's ability to establish entanglement across two \SI{25}{km}-long optical fibers independently, then to swap that entanglement efficiently to extend it over both fibers, is demonstrated. The resultant entanglement is established between telecom-wavelength photons at either end of the \SI{50}{km} channel. Finally, the system improvements to allow for repeater-node chains to establish stored entanglement over \SI{800}{\km} at Hertz rates are calculated, revealing a near-term path to distributed networks of entangled sensors, atomic clocks and quantum processors.

\end{abstract}

\maketitle

Envisioned quantum networks consist of distributed matter-based nodes, for quantum information processing and storage, that are interconnected with quantum channels for photons to establish internode entanglement \cite{Kimble2008, Wehnereaam9288}.
Such networks would enable a range of new applications \cite{Wehnereaam9288} and are desirable over distances that that span a single lab up to those required for a global quantum internet. 
The bottleneck for implementing quantum networks over long distances---the exponentially increasing probabilities of losing photons or having their information scrambled with the channel length---could be overcome by building \emph{quantum repeaters} \cite{Briegel1998, Duan2001,Sangouard2011_review, PhysRevLett.98.190503,PhysRevA.79.032325, Azuma:2015py, 7010905, Avis2022}.
With quantum repeaters, the distance between entangled end nodes in quantum network demonstrations \cite{Moehring2007, Ritter2012, Hanson2015, Ballance2020, Hanson2021,Leent:2022ek, Hofmann12, Bernien2013, Delteil2016, Stockill2017, Magnard2020, Duan2010, oxfordclock, PhysRevLett.130.050803} could be scaled up far beyond the attenuation length of the best photonic waveguides: optical fibers for telecom wavelengths. 

 \begin{figure}[t]
	\begin{center}
        \includegraphics[width=1\columnwidth]{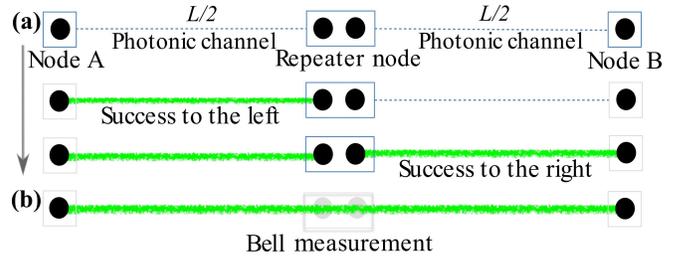}
		\caption{
		\textbf{Quantum repeater concept.} 
		a) A repeater node attempts to establish entanglement links (green lines) between its qubits (black spheres) and a qubit in each neighbouring node separately, via 
		the transmission of photons through channels on the order of the attenuation length ($L/2$). 
Whichever link succeeds first (left-hand is shown) is stored in memory 
until the remaining link succeeds. The maximum link attempt rate is inversely proportional to the photon travel time ($L/(2c)$). b) A Bell state measurement in the repeater node establishes entanglement between the neighbouring nodes.  
			}
		\label{fig:concept}
	\end{center}
\end{figure}

Following the original scheme \cite{Briegel1998}, 
 quantum repeater nodes (Figure 1) for fiber-based networks require a combination of capabilities including: 
(1) interfaces between stationary qubit registers and telecom photons and;
(2) quantum memories with storage times longer than the remote entanglement generation time.
Significant progress has been made towards developing these capabilities using ensemble-based memories \cite{Sangouard2011_review}, including with telecom interfaces \cite{Bussieres:2014zm, Pu2019, Lago-Rivera2021}. 
Memory-enhanced communication has been achieved over short distances with individual \cite{Langenfeld_Rempe_21} and ensembles \cite{Chou1316,Pu2021} of atoms, and with solid-state vacancy centres \cite{Hanson2018,Bhaskar_Lukin2020}. 
However, there are significant advantages if repeater nodes have a third capability: (3) deterministic quantum information processing \cite{Sangouard2009}.  
Without telecom-wavelength interfaces, capabilities (1)-(3) were combined in systems of vacancy centres a few meters apart and used for entanglement distillation \cite{Kalb928}, teleportation \cite{ Hermans:2022pq} and entanglement of three nodes \cite{Hanson2021}. 
Open challenges are to combine all the above capabilities in a single system and use them to demonstrate repeater-assisted entanglement distribution over tens of kilometers: 
internode distances over which repeaters could offer significant advantages for entangling matter, over repeaterless schemes.
Due to the extended photon travel times and lower success probabilities, achieving repeater-assisted entanglement distribution over tens of kilometres places far stronger demands on system performance than over a few meters. 

In this letter, a repeater node based on trapped ions is first presented that combines capabilities (1)-(3), as envisioned in \cite{Sangouard2009}, and used to demonstrate repeater-assisted entanglement distribution over a \SI{50}{km} channel. Specifically, photonic entanglement is established at the ends of two 25 km-long fiber channels with the repeater node in the middle. 
Next, the dominant cause of entanglement infidelity is identified and then the repeater node is shown to achieve a higher rate than when operated in a direct transmission configuration. 
Finally, in future, the repeater node could be duplicated and concatenated to form repeater chains;  
the performance enhancements required for such chains to establish heralded, stored entanglement between ions \SI{800}{km} apart are calculated. 

Our trapped-ion quantum repeater node is presented in Figure 2. 
The repeater node is based on two $^{40}$Ca$^+$ ions in a linear Paul trap and at the position of the waist of a near-concentric \SI{20}{mm}-long Fabry-Perot optical cavity that achieves efficient photon collection \cite{SchuppPRX2021, schuppthesis} 
at \SI{854}{nm}. Ions {separated by \SI{5.8}{\micro\meter}} are deterministically positioned at neighbouring anti-nodes of a vacuum cavity mode \footnote{See Supplemental Material including Refs.~\cite{Kirchmairthesis, PRA_2003_Jezek_likelihood, Hill97_concurrence, Pirandola2017, Scarani09,Quraishi19_swapping} at [URL will be inserted by publisher] for details on the experimental setup layout and sequence, entangled states reconstruction including unitary rotations, investigation and modelling of the ion-memory decoherence and modelling of the future systems}. 
Single photons are generated via a bichromatic cavity-mediated Raman transition (BCMRT) \cite{Stute2012, SchuppPRX2021}, driven via a  \SI{393}{nm} Raman laser beam with a \SI{1.2}{\mu }m waist at the ions. 
A Raman laser pulse on an ion in the state $\ket{S}$ (Figure 2) ideally generates the maximally-entangled state $(|D,H\rangle+|D',V\rangle)/\sqrt{2}$, where the first ket vector element describes the ion and the second describes two orthogonal polarisations of a photon emitted into the cavity. 
After the first photon has exited the cavity with high probability, the laser focus is moved to the co-trapped ion, allowing sequential generation of an ion-entangled photon from each ion. 
Each photon is converted to \SI{1550}{nm} (telecom C band), via the system of \cite{Krutyanskiy:2019cx, Krutyanskiy2017}. Next, a fiber-coupled optical switch sends photons from ion A to photonic Node A and photons from ion B to photonic Node B {(Figure 2a)}, each via separate \SI{25}{km}-long single-mode fiber spools. 
\begin{figure}[t]
	\begin{center}
                \includegraphics[width=\columnwidth]{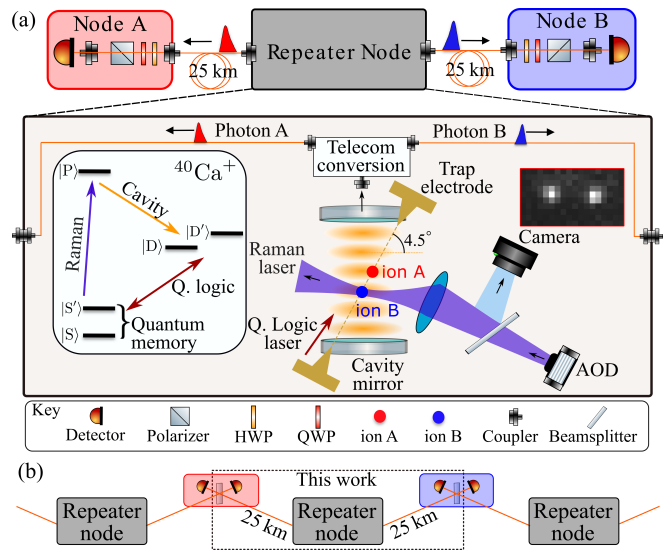}   
		\caption{
			\textbf{Trapped-ion quantum repeater node.} 
			\textbf{(a)} Experimental schematic of an elementary network segment containing one repeater node.
			Photonic nodes A and B are each sent a telecom-converted \cite{Krutyanskiy:2019cx, Krutyanskiy2017} \SI{1550}{nm} photon, entangled with a different ion, via \SI{25}{km}-long fiber spools. Ion-qubit readout is done via imaging ion fluorescence at \SI{397}{nm} on a camera. 
			Ion-qubit quantum logic {(Q. logic)} gates are done via 
			a \SI{729}{nm} laser that couples equally to the ions. AOD: Acousto-optic deflector. HWP(QWP): half-(quarter-)wave plates.
			Level scheme:  	$\ket{S}{=}\ket{4^{2} S_{1/2,m_j{=}{-}1/2}}$, $\ket{S'}{=}\ket{4^{2} S_{1/2,m_j{=}+1/2}}$, $\ket{P}{=}\ket{4^{2} P_{3/2,m_j{=}-3/2}}$, $\ket{D}{=}\ket{3^{2}D_{5/2,m_j{=}-5/2}}$, $\ket{D'}{=}\ket{3^{2}D_{5/2,m_j{=}-3/2}}$. 
			\textbf{(b)} Envisioned concatenation of the network segment in (a) into a repeater chain. Photon detection heralds remote ion entanglement \cite{Luo2009}.  
			}
		\label{fig:scheme}
	\end{center}
\end{figure}

The repeater protocol has four parts \cite{Note1}: Initialisation, Loop 1, Loop 2 and DBSM (deterministic Bell state measurement), which are now summarized.  
Initialisation prepares both ions in the state $|S\rangle$ and the ground state of the axial centre-of-mass mode. 
Next, Loop 1 begins, in which up to $29$ attempts are made to distribute ion-entangled photons to both photonic nodes. 
Each Loop-1 attempt includes two Raman pulses, one on each ion, followed by a \SI{250}{\mu s} wait time: sufficient for a photon to travel  over a \SI{25}{km} fiber and, to simulate the remote nodes being \SI{25}{km} away, for the `photon detected' signal to return.  
In the case of no detection events during all attempts in Loop 1, the protocol is restarted. 
In the case of a photon detection event at \emph{both} nodes during an attempt, Loop 1 is aborted, a \SI{729}{nm} laser pulse maps $\ket{D'}$ to $\ket{S}$ for both ions and the DBSM is performed, as described below. 
In cases where either Node A or B detect a photon during a Loop-1 attempt, Loop 1 is aborted, {729 nm laser pulses are applied to both ions, mapping $\ket{D}$/$\ket{D'}$ states to $\ket{S}$/$\ket{S'}$ states respectively}, and Loop 2 begins.  
Those pulses transfer the qubit, encoded in the ion that generated the detected photon, into the $\ket{S}$/$\ket{S'}$ spin-qubit states, which serve as a quantum memory. 
In contrast to the $\ket{D}$/$\ket{D'}$ states, the spin-qubit states have no natural spontaneous decay rate, and suffer negligible differential AC Stark shifts imparted by single photons in the cavity or by laser light periodically injected into the cavity for length stabilisation.  

In Loop 2, up to 190 attempts are made to distribute a photon, entangled with the ion not storing a qubit in memory, to the remaining remote node.  
{Periodic spin-echos are executed to extend the memory coherence time: each of which flips the $\ket{S}$/$\ket{S'}$ population and is executed using a sequence of  \SI{729}{nm} laser pulses.} 
In cases where an attempt in Loop 2 yields a photon detection event at the targeted remote node, both ion-qubits are transferred into the states $\ket{S}$ and $\ket{D}$  and the DBSM is performed, otherwise the protocol is restarted.  

Finally, the DBSM is performed in two steps \cite{Note1}. 
First a laser-driven two-qubit Molmer-Sorensen (MS) quantum logic gate \cite{PhysRevLett.82.1971, PhysRevA.62.022311, Benhelm:2008xn} is applied. 
Second, the logical state of each ion qubit is measured via fluorescence state detection captured by a digital camera. 
Observing one of the four logical states of the ion qubits heralds the projection of the polarisation of the remote telecom photons into one of four ideally orthogonal maximally-entangled states. 
Those photonic states are characterised via quantum state tomography, done by repeating the protocol for a range of polarisation analysis settings at the photonic nodes.
A fidelity over 0.5 between reconstructed and a maximally-entangled states proves that the former is entangled \cite{Sackett:2000gm}.

       \begin{figure}[h]
 \includegraphics[width=1\columnwidth]{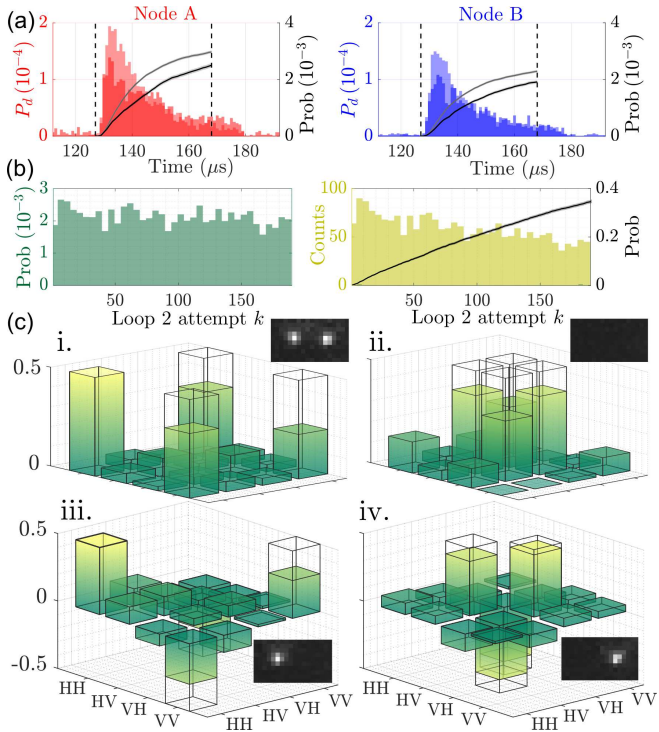}
		\caption{
			\textbf{Repeater protocol results.} \textbf{(a)} Left axes: Probability distribution ($P_d$) for single photon detection per \SI{1}{\micro\second} time bin. 
Lighter and darker coloured bars: Loop 1 and Loop 2 data, respectively.
Time origins: Raman laser pulse onset. 
Dashed black lines: windows where DBSM is performed. 
Right axes: cumulative photon detection probability for Loop 1 (lighter line) and Loop 2 (darker line).  \textbf{(b)} Photon detection probability (green bars) and counts (yellow bars) in attempt $k$ of Loop 2. 
Black line (right axis): cumulative photon detection {(count)} probability, totalling ${P_2=}\,$0.346(4).
\textbf{(c)} Coloured bars show measured real parts of the density matrix of the two single-photon polarisation qubits shared by Nodes A and B, when ion B stored a qubit in memory. Fidelities with targeted Bell states (wire mesh) are  i. $72^{+7}_{-6}~\%$ ii. $67^{+7}_{-7}~\%$. iii. $69^{+6}_{-7}~\%$ iv. $ 0.83^{+6}_{-6}~\%$. Absolute values of imaginary parts do not exceed 0.09. Inset camera images: corresponding two-ion measurement (DBSM) outcome. 
			}	\label{fig:main_entang_result}
\end{figure}

The repeater protocol was executed 44720 times over 33 minutes, during which $2,229,883$ attempts were made to establish photonic entanglement between nodes A and B, and $2053$ successes occurred ({2049 during Loop 2}). The success probability is therefore $P_s = 9.2(2)\times10^{-4}$. 
Successes correspond to asynchronous detection of a photon at each remote node, where each photon is detected within a 40 $\mu$s-wide time window containing the single photon wavepackets (Figure 3a).  
The value of $P_s$ can be compared with the expected 
probability $P^{single}_s$ to establish Node A-B entanglement without exploiting the ion memories.
We calculate $P^{single}_s = P_{A,1}\times P_{B,1}=7.2(2)\cdot 10^{-6}$ where $P_{A,1}=3.06(5)\cdot 10^{-3}$ and $P_{B,1}=2.36(4)\cdot 10^{-3}$ are the  measured probabilities for single photon detection during a Loop-1 attempt at Nodes A and B, respectively. 
The memories thus enhance the success probability by a factor $\alpha = P_s/P^{single}_s = 128$.
In principle, by performing enough Loop-2 attempts, one can approach unit probability for the second photon detection ($P_2 = 1$) yielding the maximum memory enhancement factor for our experiment of $\alpha^{max} = (P_{A,1}+P_{B,1})/(2P^{single}_s) = 375$.  
However, decoherence of the ion memories limits the number of Loop-2 attempts before stored entanglement is lost. 
In the presented experiment, 190 attempts resulted in  $P_2 = 0.346(4)$,  limiting $\alpha$ to be $\alpha = \alpha^{max}\times P_2$ (Figure 3b).

Figure 3c presents the reconstructed two-photon density matrices in the cases when 
ion B stored a qubit in memory. 
Each matrix has been rotated, from the one directly reconstructed, by the same two single-qubit rotations.  
Those local rotations, found by numerical search to maximise the average fidelities with Bell states, do not change the entanglement content of the four states, nor their inner product, and bring the measured states into the familiar Bell state form.
The Bell state fidelities of the measured states are $[72^{+7}_{-6} ,67^{+7}_{-7},69^{+6}_{-7}, 83^{+6}_{-6}]\%$.

Consequently, a different set of local rotations are used to rotate the states into the Bell form when ion A stored a qubit in memory, yielding fidelities of $[63^{+7}_{-8} ,67^{+7}_{-7},59^{+7}_{-8}, 77^{+7}_{-7}]\%$ with the corresponding Bell states. 
{Using the data underlying all 8 states}, a numerical calculation is performed to simulate the process known as feedforward, in which the photon states are further locally rotated such that the repeater always delivers a single Bell state to the remote nodes: A fidelity of $72^{+2}_{-2}\%$ is obtained \cite{Note1}.

To identify origins of protocol infidelity, the ion-photon states stored in memory are characterised in a separate experiment. Here, a modified repeater protocol is performed: up to 30 Loop 1 attempts are made to detect a photon from ion A only. In successful cases, ion-qubit A is stored in memory while a fixed number of Loop 2 attempts $k$ are made to generate a photon from ion B. 
Finally, state tomography of the joint state of ion-qubit $A$ and the detected photonic qubit `$a$'  ($\rho_{Aa}(k)$) is performed. 
That experiment is performed without fiber spools and telecom conversion for improved efficiencies, but wait times are retained allowing for twice \SI{25}{km} of photon travel.
Figure 4 shows how the fidelity of the measured $\rho_{Aa}(k)$ states, w.r.t. to the maximally-entangled state nearest to $\rho_{Aa}(0)$, evolve.  
The initial fidelity of 0.96(2) drops to 0.64(2) after 195 attempts, corresponding to a memory storage time of $t=$ \SI{64}{ms}. 
The dynamics are modelled with a single-qubit map that realises ion-qubit memory dephasing with a Gaussian temporal profile parameterised by a decoherence time $\tau$ \cite{Note1}. 
Applying the map to the initial state $\rho_{Aa}(0)$ yields modelled time-evolved states $\tilde{\rho}_{Aa}(t, \tau)$, which have fidelities  $F(\tilde{\rho}_{Aa}(t, \tau))$ with respect to the nearest maximally-entangled state to $\rho_{Aa}(0)$. 
A fit of the model to the data yields $\tau = 62\pm3$~ms (solid red line, Figure 4). 
The model is now extended to predict the two-photon states reconstructed in the full repeater protocol \cite{Note1}.  
Here, for the ion-photon states stored in memory, the states $\tilde{\rho}_{Aa}(t, \tau=62ms)$ are used for both ions. 
For the ion-photon states not stored in memory, the state $\rho_{A}(0)$ is used for both ions. 
All other parts of the repeater protocol, and feedforward, are modelled without imperfection. 
The predicted two-photon Bell state fidelities, established between the remote nodes at step $k$ of Loop 2, are plotted as a dashed black line in Figure 4. Finally, the modelled two-photon states at each attempt ($k$) are added up as a mixture, weighted by the probability with which they occurred in the repeater experiment (Figure 3b), yielding a predicted photonic Bell-state state fidelity for the repeater protocol of 0.813(7). 
Considering the experimentally obtained value of 72(2)\%, we conclude that our model captures the dominant sources of infidelity in the repeater protocol: decoherence of the ion-memories.  

\begin{figure}[t]
	\begin{center}
        \includegraphics[width=1\columnwidth]{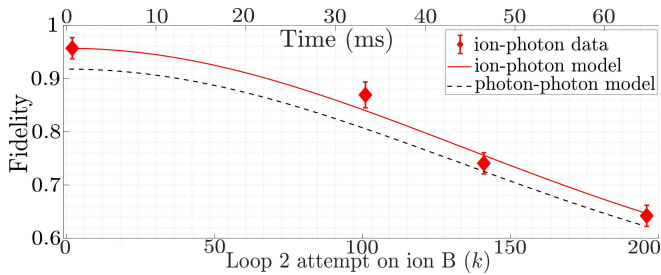}
		\caption{
			\textbf{Quantum memory performance.} 
\emph{Red diamonds}: 
Fidelity between the ion-photon state stored in ion A after $k$ photon generation attempts on neighbouring ion B, $\rho_{Aa}(k)$, and the maximally-entangled state closest to $\rho_{Aa}(0)$.
\emph{Red line}: Fit of the fidelity of the modelled ion-photon state  (see text) stored in ion A, $\tilde{\rho}_{Aa}(t)$, to the data, yielding an ion-qubit decoherence time of $\tau = $\SI{62(3)}{ms}.  
\emph{Black dashed line}: Bell-state fidelity of the modelled photon-photon state established in attempt $k$, using perfect DBSM and feedforward (see text). 
Time axis: average attempt duration. 
Error bars represent one standard deviation of uncertainty due to Poissonian photon counting statistics.
}
		\label{fig:memory}
	\end{center}
\end{figure}

The results of an independant experimental investigation into the origins of the ion-memory decoherence \cite{Note1} are now summarized. 
Raman laser pulses on the neighbouring ion during the repeater protocol 
decrease the ion-memory coherence time from $\tau=$\SI{108(1)}{ms} to \SI{59(1)}{ms}. 
A significant contribution to this decrease is made by imperfections in the spin echos caused by the initial and increasing temperature of ion string. 
The heating is caused by the ion in the Raman laser focus absorbing and spontaneously emitting \SI{393}{nm} photons. 
{No significant decoherence due to direct interactions between an ion's memory qubit and the Raman laser  pulses on the neighbouring ion was found.}

In our last experiment, performance in repeater configuration is compared with that in a direct transmission configuration. 
For direct transmission, the fiber spools are joined into a \SI{50}{km} channel. 
Furthermore: only the initialisation and Loop 1 of the repeater protocol are executed; both ion-entangled photons are directed to Node B and; the detection of any one photon counts as a success. 
The wait time after photon generation is set to \SI{500}{\mu s} to allow for twice the \SI{50}{km} travel time. 
In both configurations, attempt rates are predominantly limited by the photon travel and signal return times, and polarization analysis is removed to improve overall efficiency. 
Having previously observed maximal ion-photon entanglement over \SI{50}{km} with a fidelity of 0.86$\pm{0.03}$ in direct configuration with a single ion \cite{Krutyanskiy:2019cx}, we focus here on measuring the success rates. 
The results are absolute success rates of \SI{5.9}{Hz} and \SI{3.8}{Hz} for the repeater and direct transmission configurations, respectively. 
After excluding slight differences in initialisation and dead times \cite{Note1} we obtain success rates of \SI{9.2}{Hz} and \SI{6.7}{Hz}, where the repeater is still faster. 
Analytic expressions for the performance requirements for a repeater node to beat itself in direct transmission are summarised in \cite{Note1}.

{
We now present parameter combinations for future enhanced version of our repeater nodes that would enable a significant advance in the state of the art \cite{Leent:2022ek} for entanglement distribution between remote matter. We consider the chains of enhanced repeater nodes (Figure 2b) and model of \cite{Sangouard2009} to calculate the heralded entanglement rate between ions in the end nodes.  

The key model parameter is $P^0_{link}=0.21$: the combined telecom photon generation and detection probability for each node excluding channel (fiber) losses. 
For a chain of two such nodes, with a \SI{50}{km} telecom fiber between them, 
the predicted average time to establish heralded remote ion-ion entanglement is \SI{0.07}{s}.
Furthermore, by chaining 17 of those enhanced nodes across an \SI{800}{km} fiber, 
the predicted average time to establish entanglement between one ion in each end node is \SI{0.7}{s}.
Despite the channel loss probability increasing by 13 orders of magnitude (50 to \SI{800}{km}), and commensurate rate reduction for direct transmission schemes, the repeater-assisted entanglement distribution rate is predicted to drop by only a factor of 10.}

{To quantify the requirements for entanglement to be present across the 800 km chain, we introduce a simple model for the established Bell state fidelity \cite{Note1}. The model does not consider memory decoherence, which is valid under the condition that the memory coherence time in each node is much longer than the average entanglement distribution time between the end nodes, that is $\tau\,>>\,$\SI{0.7}{s}. 
The model parameters are: $F_{0}$, the ion-photon Bell state fidelity; $F_{swap}^{ions}$, the DBSM fidelity and; $F^{photon}_{swap}$, a non-deterministic photonic Bell state measurement fidelity.  For the case where all three aforementioned fidelities are 0.99, the model predicts a Bell state fidelity of 0.61, which is over the 0.5 threshold. }

The required enhanced ion-node parameters are ambitious but feasible. 
The current repeater node achieves $[{P^0_{link}=}\, 0.018(1), {\tau=}\, 62(3)$ ms$, {F_{0}=} \,0.96(2), {F_{swap}^{ions}=} \,0.95(2), {F^{photon}_{swap}=} \,0.69(2)]$, where the last value is taken from the two-photon interference contrast at telecom wavelengths measured in \cite{Krutyanskiy:2020pra}. The required improvements in $F_0$ and $F_{swap}^{ions}$ are minor, while those in $P^0_{link}$ may already be possible by combining the single ion-photon collection efficiencies of \cite{SchuppPRX2021} and the telecom conversion efficiency of \cite{Bockthesis}. 
Memory times of tens of seconds in the repeater node should be possible by combining decoherence-free subspaces \cite{Zwerger2017, Haffner:2005ls} with sympathetic cooling \cite{PhysRevLett.98.220801,PhysRevA.79.050305} {or directly using other isotopes \cite{Drmota2022}}. 
Achieving the required $F^{photon}_{swap}$, whilst maintaining $P^0_{link}=0.21$, may be achieved by combing the methods of \cite{SchuppPRX2021, PhysRevA.102.032616} with a smaller mode-volume cavity {and careful attention to minimise birefringence \cite{Kassa20}}.

A long-distance quantum repeater node based on trapped ions with an optical cavity was demonstrated, as envisioned in \cite{Sangouard2009}. 
Two remote trapped ions have previously been entangled over a few meters \cite{Moehring2007,  Ballance2020,Duan2010, oxfordclock} and, using the ion-cavity system of this work, over \SI{230}{m} \cite{PhysRevLett.130.050803}. 
The results in this work, in combination with those in \cite{PhysRevLett.130.050803, Krutyanskiy:2020pra}, demonstrate all key capabilities of a long-distance quantum repeater platform in a single system. In future, more qubits could be directly integrated into the node \cite{Sangouard2009} allowing for quantum error correction and multimoding  \cite{PhysRevLett.98.190503}.

Datasets are available online~\footnote{Will be provided upon acceptance}

\begin{acknowledgments}
This work was financially supported by the START prize of the Austrian FWF project Y 849-N20, 
the Austrian FWF Standalone project QMAP with project number P 34055,  
the Institute for Quantum Optics and Quantum Information (IQOQI) of the Austrian Academy Of Sciences (OeAW) and the European Union’s Horizon 2020 research and innovation program under grant agreement No 820445 and project name ‘Quantum Internet Alliance’. 
We acknowledge funding for V. Krutyanskiy by the Erwin Schr\"odinger Center for Quantum Science \& Technology (ESQ) Discovery Programme, 
for N.S. by the Commissariat \`a l'Energie Atomique et aux Energies Alternatives (CEA),
and for B.P.L. by the CIFAR Quantum Information Science Program of Canada.

Experimental data taking was done by V.Kru, M.M., V.Krc and M.C.
Development of the experimental setup was done by V.Kru, M.C., M.M, J.B., J.S. and B.P.L.
Data analysis and interpretation was done by V.Kru, M.C., M.M. J.B., and B.P.L 
Modelling was done by V.Kru, J.B., M.C. and N.S.  
The manuscript was written by B.P.L. and V. Kru, all authors provided detailed comments. 
The project was conceived and supervised by B.P.L.
\end{acknowledgments}

\bibliography{repeater_bib.bib}

\end{document}


\title{A telecom-wavelength quantum repeater node based on a trapped-ion processor: 
Supplemental material }

\author{V.~Krutyanskiy}
\affiliation{Institut f\"ur Experimentalphysik, Universit\"at Innsbruck, Technikerstr. 25, 6020 Innsbruck, Austria}

\affiliation{Institut f\"ur Quantenoptik und Quanteninformation, Osterreichische Akademie der Wissenschaften, Technikerstr. 21a, 6020 Innsbruck, Austria}

\author{M.~Canteri}
\affiliation{Institut f\"ur Quantenoptik und Quanteninformation, Osterreichische Akademie der Wissenschaften, Technikerstr. 21a, 6020 Innsbruck, Austria}
\affiliation{Institut f\"ur Experimentalphysik, Universit\"at Innsbruck, Technikerstr. 25, 6020 Innsbruck, Austria}

\author{M.~Meraner}
\affiliation{Institut f\"ur Quantenoptik und Quanteninformation, Osterreichische Akademie der Wissenschaften, Technikerstr. 21a, 6020 Innsbruck, Austria}
\affiliation{Institut f\"ur Experimentalphysik, Universit\"at Innsbruck, Technikerstr. 25, 6020 Innsbruck, Austria}

\author{J.~Bate}
\affiliation{Institut f\"ur Experimentalphysik, Universit\"at Innsbruck, Technikerstr. 25, 6020 Innsbruck, Austria}

\author{V.~Krcmarsky}
\affiliation{Institut f\"ur Quantenoptik und Quanteninformation, Osterreichische Akademie der Wissenschaften, Technikerstr. 21a, 6020 Innsbruck, Austria}
\affiliation{Institut f\"ur Experimentalphysik, Universit\"at Innsbruck, Technikerstr. 25, 6020 Innsbruck, Austria}

\author{J.~Schupp}
\affiliation{Institut f\"ur Quantenoptik und Quanteninformation, Osterreichische Akademie der Wissenschaften, Technikerstr. 21a, 6020 Innsbruck, Austria}
\affiliation{Institut f\"ur Experimentalphysik, Universit\"at Innsbruck, Technikerstr. 25, 6020 Innsbruck, Austria}

\author{N.~Sangouard}
\affiliation{Institut de Physique Th\'eorique, Universit\'e Paris-Saclay, CEA, CNRS, 91191 Gif-sur-Yvette, France}

\author{B.~P.~Lanyon}
\email[Correspondence should be send to ]{ ben.lanyon@uibk.ac.at}
\affiliation{Institut f\"ur Experimentalphysik, Universit\"at Innsbruck, Technikerstr. 25, 6020 Innsbruck, Austria}
\affiliation{Institut f\"ur Quantenoptik und Quanteninformation, Osterreichische Akademie der Wissenschaften, Technikerstr. 21a, 6020 Innsbruck, Austria}

\date{\today}

\maketitle

\tableofcontents

\section{Experimental setup and protocol}
\label{sec:setup}

\subsection{Ion trap and photon conversion}
\label{sec:trapandconversion}

For details on the construction and design of our cavity-integrated ion trap system see \cite{schuppthesis,SchuppPRX2021}. 
In summary, our macroscopic linear Paul blade ion trap hangs via rigid attachment from the top flange of a vacuum chamber and is orientated such that the ion string axis is in the vertical direction. 
An in-vacuum optical cavity around the ion trap, with axis a few degrees off horizontal, is mounted on the bottom flange via nanopositioning stages.
The \SI{19.906(3)}{mm}-long near-concentric cavity has a $12.31(8) \mu m$ waist at the point of the ions and 10 mm ion-mirror separation.
{ For more details on the cavity characterization including the mirrors' losses, birefringience and ion-cavity coupling see  \cite{schuppthesis}.} 
Ions are loaded using a resistively-heated oven and a two-photon ionisation process involving lasers at \SI{422}{nm} and \SI{375}{nm}. 
For the experiments reported in this paper, the axial center-of-mass (COM) trap frequency was  0.963(5) MHz, while the radial COM frequencies were 2.18(1) and \SI{2.13(1)}{MHz}.\\

\noindent\textbf{Geometry}.
Consider a Cartesian coordinate system with three orthogonal directions: $x$, $y$ and $z$.
The $z$ direction points along the axial center of mass motion of the ion string, defined by the line connecting the vertically-orientated DC endcap electrodes of the linear-Paul ion trap. The $z$ direction is therefore the one along which the ion string lines up. 
The $xz$ plane is defined as the plane containing both the $z$ axis and the cavity axis. The cavity axis subtends an angle with respect to the $x$ axis of {\SI{4.5(3)}{\degree}}.

The atomic quantisation axis is chosen to be parallel to the axis of an applied static magnetic field. This magnetic-field axis is set to subtend an angle of \SI{45}{\degree} with respect to the $z$ axis and to be perpendicular to the cavity axis. It is likely that it is a few degrees off from perpendicular. A magnetic field of \SI{4.22892(2)}{G} is set by permanent magnets attached to the outside of the vacuum chamber. Field strengths are measured via Ramsey spectroscopy of a single ion.
Since the cavity is near perpendicular to the quantisation axis, \SI{854}{nm} photons emitted into the cavity on transitions where the magnetic quantum, $m$, number changes by 1 (0) are projected onto vertical (horizontal) polarization. \\

\noindent\textbf{Photon conversion}.
Details on the polarization-preserving photon frequency converter are given in \cite{Krutyanskiy:2019cx,Krutyanskiy2017}, in particular, see fig. 5 of \cite{Krutyanskiy2017} . 
The differences between the setup of \cite{Krutyanskiy2017} and the present one are now described. In the present setup: 1. The aspheric lenses (AS) are now anti-reflection coated 2. A 1m SMF-28 fiber has been added before the holographic grating filter. 3. The PBS in fig. 5 of \cite{Krutyanskiy2017} is removed. 
The maximum fiber to fiber photon conversion efficiency observed in the present work was 27\% (see also \refsec{sec:efficiency}).
The single photons at telecom wavelength are detected with superconducting nanowire detectors with efficiencies of 74\% and 75\% at \SI{1550}{nm} and that both have free running dark count rates of $<0.6(1) s^{-1}$, as measured by the manufacturer on installation. 
The detectors were manufactured by `LLC Scontel'.

\subsection{Single-ion focused photon generation method}
\label{Photon_generation_method}

\begin{figure}[h!]
	\vspace{0mm}
	\begin{center}
        \includegraphics[width=0.99\columnwidth]{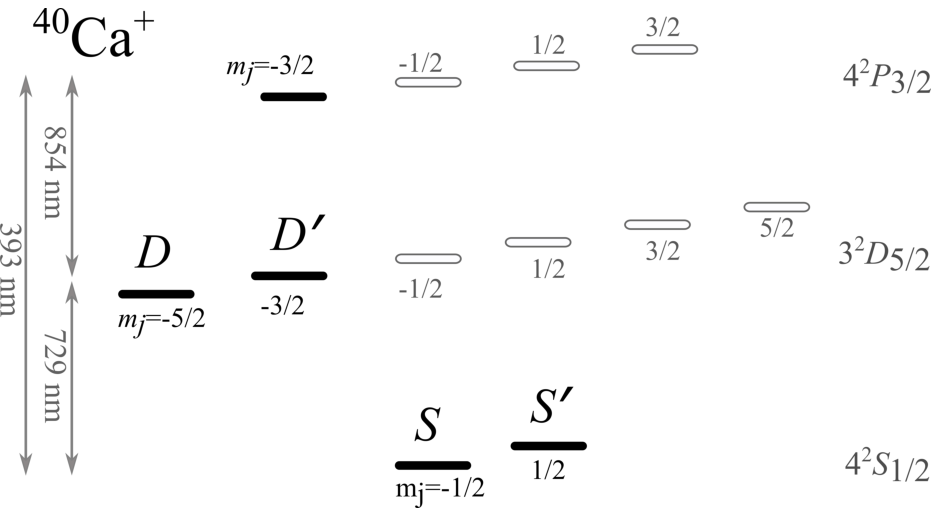}
		\vspace{-5mm}
		\caption{
			\textbf{Atomic energy level diagram of  $^{40}\mathrm{Ca}^+$}. Three fine structure manifolds are shown, and the Zeeman splitting within each of those manifolds due to the applied magnetic field of a few Gauss. The five bold levels are discussed in this supplemental material. The shorthand notion for those five levels is $\ket{S}{=}\ket{4^{2} S_{1/2,m_j{=}{-}1/2}}$, $\ket{S'}{=}\ket{4^{2} S_{1/2,m_j{=}+1/2}}$, $\ket{P}{=}\ket{4^{2} P_{3/2,m_j{=}-3/2}}$, $\ket{D}{=}\ket{3^{2}D_{5/2,m_j{=}-5/2}}$, $\ket{D'}{=}\ket{3^{2}D_{5/2,m_j{=}-3/2}}$. }
		\label{figsup:Levels}
		\vspace{2mm}
	\end{center}
\end{figure}

\noindent\textbf{Overview.}
As summarized in the main text, photons are generated from the ions via a bichromatic cavity-mediated Raman transition (BCMRT) driven by a single-ion-focused \SI{393}{nm} Raman laser  beam. An ion in the focus of the laser beam emits a photon into the \SI{854}{nm} vacuum cavity mode. The polarization of that photon is entangled with the final electronic state of the ion that emitted the photon. 

The BCMRT has been realised several times before with trapped ions e.g., \cite{Stute2012, SchuppPRX2021}.  The present work is the first realisation of a single-ion addressed version and more generally: a technique that allows the extraction of sequential photons each entangled with different ion-qubits in a string. 
Besides the efficient photon collection enabled by the cavity (see \ref{sec:WP_simulation}), a second key feature of the technique 
is that the photon generation process on the ion of choice is strongly decoupled from the states of co-trapped ions. Consequently, the electronic states of the co-trapped ions are suitable for storing quantum information (c.f. Figure 4 of the main text) or are ready to generate subsequent photons without reinitialisation (c.f. Loop 2 of the repeater protocol). Furthermore, tomographically-reconstructed ion-photon states show that each photon is strongly entangled with the ion selected to emit it and display no entanglement with ions not selected to emit it. The origins of this feature of our addressed cavity-mediated photon generation process --- strong decoupling between the photon generation process and the states of a co-trapped ion in the cavity standing wave---are now shortly summarised. Theoretical predictions and experimental verification of the fidelity limits to the technique are left for future work. 

Consider a two ion string, with each ion positioned near a maximum (anti-node) of the vacuum cavity standing wave. Consider the ion not in the focus of a Raman laser beam: its coupling strength in the BCMRT is reduced not only due to the tight focusing of the laser intensity but also because that nearby laser light is off-resonant. Specifically, laser frequencies are set to resonantly drive the Raman transition of the ion in focus, which is AC Stark-Shifted by 0.7 MHz compared to the other ion: larger than the transition linewidth for either ion. Consequently, the probability per attempt that the non-addressed ion either generates its own cavity photon, or absorbs the emitted cavity photon of its neighbour, is vanishingly small.  Second, differential AC-Stark shifts introduced by single cavity photons on a qubit encoded into the D/D' states are negligible in our ion-cavity parameter regime. Those single-photon shifts are on the order of a few kHz \cite{PhysRevLett.122.153603} while each photon spends only 1.14(2) $\mu$s on average inside the cavity (cavity ring-down time \cite{SchuppPRX2021}).  \\

\noindent\textbf{Experimental details.} 
Details on the optical setup of the single-ion focused Raman laser are now provided. More information can be found in the Master thesis of group member Marco Canteri \cite{Marcothesis}. In summary, a \SI{393}{nm} laser beam is focused onto the ion string via a custom-made multi-lens objective whose front surface lies approximately 59 mm from the ion string. The objective has a numerical aperture of 0.29 
and is outside the vacuum chamber, inside an inverted viewport\footnote{the objective serves to both focus the \SI{393}{nm} Raman laser onto the ion string as well as to collect 397nm flourescence photons from the ions and image them onto a digital camera, as described in \cite{Marcothesis}}. 
The optical axis of the objective is set to be parallel to $y$ axis (\refsec{sec:trapandconversion}) and is therefore: perpendicular to the ion string axis ($z$ direction); within a few degrees of perpendicular to the cavity axis and; at 45 degrees to the quantisation axis set by the principle magnetic field ($y$/$x$ plane).
The polarization of the \SI{393}{nm} beam is set to be linear and perpendicular to the quantisation axis and therefore only drives atomic dipole transitions that change the magnetic quantum number ($m_j$) by plus or minus one. As previously stated, since the cavity is approximately  perpendicular to the quantisation axis, \SI{854}{nm} photons emitted into the cavity on transitions where the magnetic quantum, $m$, number changes by 1 or 0 are projected onto vertical or horizontal polarization, respectively.

An acousto-optic deflector\footnote{Gooch \& Housego 4120-3} (AOD) in the \SI{393}{nm} beam path allows the focal point of the beam to be moved along the ion string axis over a range of 160 $\mu$m in 8 $\mu$s \cite{Marcothesis}.  A double-pass acousto-optic modulator (AOM), before the AOD as the light flies, corrects for the frequency shift imparted by the AOD such that shifting the position of the ion focus does not cause a frequency shift of the laser at the point of the ion. 
At the ions, the \SI{393}{nm} beam is measured to have a waist of 1.2(2) $\mu$m ($e^{-1}$ intensity radius) which can be compared with the ion-ion separation of 5.8 
$\mu$m. 

At the point of the ion, the Raman drive laser contains two frequencies: one detuned by $\Delta$ from the $|S\rangle=|4^2\text{S}_{1/2}, m_j = -1/2\rangle$ to $|4^2\text{P}_{3/2}, m_j = -3/2\rangle$  transition, and the other detuned by $\Delta+\delta$ from the same transition. 
The detuning $\Delta/2\pi = -377(5)$ MHz, as measured by a wavemeter.
The detuning $\delta = -7.10466(3)$ MHz is the Zeeman splitting between the $|D\rangle=|3^2\text{D}_{5/2}, m_j = -5/2\rangle$ and neighbouring state $|D'\rangle=|3^2\text{D}_{5/2}, m_j = -3/2\rangle$, as measured via Ramsey spectroscopy at 729 nm in the absence of the Raman laser pulse. 
The two tones in the Raman laser are generated by applying two RF tones (differing by 7.104663 MHz) to a single pass AOM before the AOD. 

The two tones in the Raman drive laser simultaneously drive two cavity-mediated Raman processes, for the ion in the focus.
One process generates a horizontally polarised (H) cavity photon and the other a vertically polarised (V) cavity photon. 
In the case where the two processes occur with equal probability, the initial state $\ket{S}=|4^2\text{S}_{1/2}, m_j = -1/2\rangle$ (with no photons in the cavity) is ideally transferred to the final,
maximally entangled state $(|D,H\rangle+e^{i\beta}|D',V\rangle)/\sqrt{2}$ 
where the phase $\beta$ is set by the relative phase
of the two laser fields.
The individual powers of the two drive-laser fields are varied until the probability for detecting photons is balanced in the H/V polarization basis, to within statistical uncertainty. 

The effective Rabi frequency of the bichromatic field is required for the numerical simulations of the photon generation process in \refsec{sec:WP_simulation}. For that, we measure the AC Stark shift introduced by the laser field on the bichromatic Raman transition, which can be related to the effective Rabi frequency given known detunings. To determine that shift, we first determine the center frequency of the Raman transition, corresponding to the value at which the photon generation probability is highest. Next, we repeat the measurement for vanishingly small laser power. The difference in the two center frequencies is found to be 0.70(2) MHz when focusing the Raman beam on ion A, which represents an estimate for the AC Stark shift introduced by the full power beam. The same experiment on ion B yielded a shift of 0.77(2) MHz. 
The  value of 0.7 MHz is used for the simulations in \refsec{sec:WP_simulation}.

\subsection{Positioning the ions in the cavity}
The methods for deterministically positioning the ions into the vacuum cavity standing wave are now summarised, followed by estimates of the ion-cavity coupling strengths. 
The linear ion string is located at the position of the waist of the cavity. 
The axial COM trap frequency of 963 kHz  yields a calculated ion-ion separation of  $d=5.8 \mu$m. 
When projected onto the cavity axis, the ion's separation is given by $d\cos{\theta}=455$ nm, with $\theta=85.5^\circ$. The distance between neighbouring antinodes of the vacuum cavity standing wave is $\lambda /2=427 $nm, where $\lambda=854$nm . Therefore, for the employed axial confinement, both ions cannot be placed simultaneously at two antinodes, where they each would achieve maximum ion-cavity coupling strength. 
{The axial confinement used was limited from above by specifications of a capacitor used as a part of the circuit that provides axially confining voltage.}

The maximum ion-cavity coupling strength that can be achieved for a single ion trapped in our system is $g^{max} = 2\pi\times 1.53(1)$ MHz \cite{SchuppPRX2021}. 
By calculating the vacuum electric field distribution in our cavity given knowledge of its geometry \cite{SchuppPRX2021}, we can subsequently calculate the ion-cavity coupling strengths for arbitrary positions of our two-ion string. 
Consider the case in which the two ions are placed equal distance ($5.8/2 ~\mu$m) away from, and either side of, the cavity axis.  In the particular case in which one of those ions is positioned precisely at a vacuum standing wave maxima (anti-node), then that optimised ion has a calculated ion-cavity coupling strength of $0.95\times g^{max}$. Simultaneously, the other ion will have a lower ion-cavity coupling strength of $0.92\times g^{max}$ . 
We do not aim for that particular case, but instead for a configuration in which the ion-cavity coupling is equal for both ions. Specifically, the ions are positioned with respect to the vacuum cavity standing wave according to the following procedure.

First, the focus of the Raman beam is set onto the ion of choice. Second, cavity photons are generated continuously from that ion via the combination of a continuous Raman laser light at \SI{393}{nm} and repumping laser light at \SI{854}{nm}. 
Third, the cavity position is translated along its axis on the nanometer scale until a position of maximum detected photon rate is obtained. 
Fourth, the Raman beam focus is set onto the other ion and the cavity photons are generated and detected with a rate $R$ counts per second. 
Fifth and finally, the cavity position is translated along its axis again until the observed detection rate of the photons is approximately half way between $R$ and the maximum achievable for this ion.
If step five of the positioning procedure is done perfectly, both ions are expected to be coupled equally with a strength of $0.935\times g^{max}$: that value is used for further calculations in \refsec{sec:WP_simulation}.

\subsection{Telecom fiber switch} 
To switch the telecom photons, we used a commercially available polarization-preserving switch from Photonwares Model NSSW 1x2.  The switch has single-mode fiber at each interface. Directly after the conversion setup, the telecom photons are collected into a fiber and sent to the single SMF-28 input fiber of the switch. 
The input can be switched using a TTL signal to either of two SMF-28 fiber outputs in \SI{0.3}{\mu s}. The total in-out efficiencies are 86(3)\%  and 82(3)\% and corresponding cross-talks are 0.5(3)\% and 1.7(3)\%, respectively. 
The polarization dependence of the device efficiency is below 1.8(3)\%. The aforementioned performances of the switch were characterized by us using a standard power meter and a manual fiber polarization controller. \\

\subsection{The repeater protocol: laser-pulse sequence}
\label{sec:sequence}
Figure 3 of the main text presents results from our repeater protocol: entanglement of telecom photons detected by Alice  (at Node A) and Bob (at Node B). The repeater protocol involves an experimental sequence of laser-driven operations on the ions that can be divided into four parts: 1. Initialisation; 2. Loop 1; 3. Loop 2 and 4. DBSM (Deterministic Bell state measurement). The details of each part are now given. In what follows the waiting times below 10$~\mu$s between operations are not specified. \\

\noindent \textbf{Initialisation.} 
Initialisation begins with 4 ms of Doppler cooling using a combination of 397 nm, 866 nm and 854 nm beams. The latter is used to empty (`repump') the $D_{5/2}$ manifold. 
Next a 150 $\mu$s pulse of monochromatic 393 nm Raman laser pulse is applied---used as part of a sample-and-hold system to stabilize the laser power---followed by another 854 nm repumping pulse and \SI{50}{\mu s} of Doppler cooling. 
The ions are then optically-pumped into the state $\ket{S}{=}\ket{4^{2} S_{1/2,m_j{=}{-}1/2}}$ via a 20 $\mu$s pulse of circularly-polarised 397-nm light pointing along the quantisation axis set by the applied magnetic field, combined with a simultaneous 866 nm repumper pulse. 
The axial COM mode is then prepared close to the ground state (0.043(4) phonons) by applying 3 ms of sideband-resolved ground state cooling on the 729 nm quadrupole transition. Optical pumping at 397nm is then performed again.  

Until this point, the optical cavity surrounding the ion has been filled with \SI{806}{nm} laser light, used for locking the cavity length. Now, that \SI{806}{nm} light is switched off and the piezo-actuators used to control the cavity’s length are held at constant voltage via a sample and hold system.
After the complete initialisation sequence, both ion qubits are prepared in the electronic state $\ket{S}$, the axial COM is prepared close to the motional ground state and the other 5 motional modes are at Doppler cooled temperatures. For more detail on the motional state of the ion string at this point, see section \refsec{sec:temp}.\\

\noindent \textbf{Loop 1.}
In loop 1, up to 29 attempts are made to distribute ion-entangled photons to both Alice and Bob. {The number of attempts in Loop 1 is limited to 29 by the experimental control hardware\footnote{ {Our central experimental control system was not able to produce more photon generation attempts in Loop 1, given the number in Loop 2. That limit comes from a combination of the way in which that system is programmed, the way it communicates with other devices and its finite memory.  More Loop 1 photon generation attempts were only possible by reducing the number in Loop 2. We chose to split the available number of photon generation attempts approximately equally between the two loops (on average), based on our estimates before carrying out the experiment.}}.}
The sequence that is looped here consists of the following operations. 

\begin{itemize}
\item A 20 $\mu$s repumping pulse at 854 nm starts simultaneously with a 25 $\mu$s \SI{866}{nm} pulse which empty the $D_{5/2}$ and $D_{3/2}$ manifolds, respectively, followed by a 20 $\mu$s  optical pumping pulse (397 nm) that starts simultaneously with a 40 $\mu$s  866 nm pulse to initialise both ions into the state $\ket{S}$. 
\item A 50 $\mu$s photon generation Raman pulse (henceforth refered to as a `Raman pulse') is applied to ion A.
\item A 10 $\mu$s wait time to allow for both switching of the Raman beam focus to ion B and, after the first photon is sent to node A, the output port of the telecom fiber to be switched to Node B. 
\item A 50$\mu$s Raman pulse is applied to ion B.
\item A 250 $\mu$s wait time. This time is long enough for both photons to travel to their respective nodes via \SI{25}{km} of fiber and any successful detection signals to return from Alice and Bob over that distance. Specifically, if a photon is successfully detected at a node during the target time windows (Figure 3a), then that node sends a TTL `success' signal to the ion control system. 250 $\mu$s allows for light travel twice over \SI{25}{km} of fiber. In the experiment the success signals were sent via a few meters of BNC cables. 
\end{itemize}

What happens next depends on whether nodes A and B reported successful detection events in Loop 1 or not (\emph{success signals}). 
In the case of no such success signal, the sequence is repeated, up to a maximum of 29 times.
After every 8th execution of the sequence, 1 ms of side-band cooling on the axial COM mode is applied. 
In the case of a success signal from \emph{both} nodes, the loop is aborted, a $12~\mu$s 729 nm laser  $\pi$-pulse maps $\ket{D'}$ to $\ket{S'}$ (Figure \ref{figsup:Levels}) for both ions and the Bell-state measurement sequence is performed, as described below.
In cases of a success signal from one node, Loop 1 is aborted and Loop 2 begins. We refer to Loop 2 as the memory loop, since a qubit is stored in one of the ion-memories during its execution. \\

\noindent \textbf{Loop 2.}
This sequence of operations takes place always in the case where one ion has already successfully distributed its photon (in Loop 1), but the other has not. The sequence here begins by transferring the qubit state of the ion that was successful from the $\ket{D}$/$\ket{D'}$ states into the $\ket{S}$/$\ket{S'}$ memory states. This transfer is done by two sequential \SI{729}{nm} $\pi$-pulses (13 and 12 $\mu$s in length). Those pulses operate on both ions and therefore serve to prepare the unsuccessful ion qubit into a mixture of the $\ket{S}$ and $\ket{S'}$ states and are done only ever once, at the beginning of the Loop 2 sequence. 

Next, the following sequence begins, which is looped up to 190 times in the case of no success signal: 

\begin{itemize}
\item A 20 $\mu$s repumping pulse at 854 nm starts simultaneously with a 25 $\mu$s \SI{866}{nm} pulse that empties the $D_{5/2}$ and $D_{3/2}$ manifolds. Optical pumping to the state $\ket{S}$ is not performed as our method would also pump the memory qubit.
\item {In cases where \SI{806}{nm} laser light is now being sent into the cavity, that light it is now turned off.} 
\item A 50 $\mu$s Raman pulse is applied to the ion that does not contain a qubit stored in memory.
\item A 250 $\mu$s wait time to allow for photon travel time and a TTL `success' signal to be received (or not). {During that wait time, 56 $\mu$s after the Raman pulse finishes, the \SI{806}{nm} laser light used for the cavity length stabilization is turned on}. 
\end{itemize}

In the case where no success signal is received from the target node, the above bullet-pointed Loop 2 sequence is repeated, up to a maximum of 190 times.
After every 5th repetition of that sequence, a spin echo is performed corresponding to a sequence of three \SI{729}{nm} $\pi$-pulses applied to both ions, as described in section \refsec{sec:spinecho}. 
The \SI{806}{nm} laser light used for the cavity length stabilization is switched off for the time of those 729 nm pulses. 
In the case of no success signal from the target node during any attempts in Loop 2, the entire repeater protocol is restarted. 
In the case of a success signal from the target node during an attempt in a single execution of the sequence in Loop 2, the loop is aborted {and the cavity stabiliziation light is turned off}. If an odd number of spin-echo sequences has been executed by this time, one more is executed using addressed single qubit manipulation, as described in \refsec{sec:recombination}. As a result, the total number of spin echos  is always even. 

At this point, the sequence moves into the DBSM part.\\

\noindent \textbf{DBSM (Deterministic Bell state measurement).}
The DBSM sequence begins with single ion-qubit manipulation as described in \refsec{sec:recombination}. After that manipulation, both ion-qubits become encoded in the $\ket{D}/\ket{D'}$ states and the  sequence proceeds as follows:  

\begin{itemize}
\item  A $12~\mu$s \SI{729}{nm} $\pi$-pulse on the $\ket{D'}$ to $\ket{S}$ transition is applied to both ions. Both ion-qubits are now encoded in the $\ket{S}/\ket{D}$ states. 
\item A 107 $\mu$s long \SI{729}{nm}-laser-driven Molmer-Sorensen (MS) gate is applied to both ions, as described in section \refsec{DBSM}.
\item {The cavity length stabilization laser light is turned on.} 
\item A \SI{2}{ms}-long ion-qubit state readout operation is performed via electron shelving. Here, \SI{397}{nm} and \SI{866}{nm} laser light are shone onto the ion string while any fluorescence photon scattering at \SI{397}{nm} is imaged on an EMCCD camera. The camera resolves the fluorescence state of the ions individually. \\
\end{itemize}

The total of the times written above, including the pulses in the single ion-qubit manipulation, is \SI{2143}{\mu}s.
Due to some additional sequence delays in the programming, the total time taken for the DBSM operation was \SI{2157}{\mu}s. This time is used as $T_{swap}$ in  \refsec{sec:compare_direct}\\

\noindent The repeater protocol is now complete.

\subsubsection{Details on addressed single ion-qubit manipulation}
\label{sec:recombination}

In the cases where photon distribution has just been successful in Loop 2, one ion-qubit is encoded in the $\ket{S}$/$\ket{S'}$ memory states of `ion 1' whereas the other ion-qubit---having just successfully generated a photon in Loop 2---is encoded in the $\ket{D}$/$\ket{D'}$ states of `ion 2'. 
The method employed for performing an entangling logic gate requires both ion-qubits to be encoded in the same states. We now describe the `recombination' process used to bring both ion-qubits to the same states using single-ion focused operations.

Ion 2 (with qubit encoded in the $D/D'$) is illuminated with the focused Raman laser beam with a single frequency detuned by $\Delta$ from the $|4^2\text{S}_{1/2}, m_j = -1/2\rangle$ to $|4^2\text{P}_{3/2}, m_j = -3/2\rangle$ transition, as defined in \refsec{Photon_generation_method}. 
Simultaneously, a sequence of two \SI{729}{nm} $\pi$-pulses, each taking \SI{12}{\mu}s, are implemented which exchange populations between states on the transitions that are resonant with them. The first $\pi$-pulse is resonant with the  $\ket{S}$ to $\ket{D}$ transition in the ion not in the Raman beam focus. The second $\pi$-pulse is resonant with the $\ket{S'}$ to $\ket{D'}$ transition in the ion not in the Raman beam focus. The \SI{729}{nm} laser field for these operations is broadly focused, illuminating both ions. However, the $\ket{D}$/$\ket{D'}$ qubit encoded in the ion 2 (in the Raman laser focus) remains unaffected, since the \SI{393}{nm} light shifts the \SI{729}{nm} transitions in that ion far off resonant via the AC-Stark effect, such that the 729 nm pulses have no significant effect.  An AC Stark shift of around 300  kHz is chosen, by choice of Raman beam intensity, which avoids bringing sideband transitions on resonance.
As a consequence, the \SI{729}{nm} $\pi-$pulses affect only ion 1, mapping its $\ket{S}$/$\ket{S'}$ qubit to $\ket{D}$/$\ket{D'}$.
Both ion-qubits are thus finally encoded in the $\ket{D}$/$\ket{D'}$ states and can be further manipulated with global 729 nm pulses when the Raman beam is turned off.

In a calibration experiment for addressed $\pi-$pulses described above  we find $<1\%~\pi$-pulse error for the ion that is not in the focus of the Raman beam. For the ion that is in the focus of the Raman beam we measured 1(1)\% and 4(2)\% probability of unwanted transition.

\subsubsection{Details on spin echo pulses}
\label{sec:spinecho}

A spin echo is performed after every 5th attempt in the Loop 2 of the repeater protocol. 
In cases where all 190 attempts are executed in Loop 2, a total of 38 spin echos are implemented. 
The total number of spin echos implemented is always even, as described in the main text. 

Each spin echo corresponds to a sequence of three sequential $12~\mu$s $\pi$-pulses using the \SI{729}{nm} laser beam. The first $\pi$-pulse exchanges electron population between the states $\ket{S}$ and $\ket{D'}$. The second $\pi$-pulse exchanges electron population between the states $\ket{D'}$ and $\ket{S'}$. The third $\pi$-pulse is the same as the first one.  The phases of the first and last pulses 
are the same, while the second pulse has a phase shift of $\pi$ with respect to the others. 
In the ideal case, this composite pulse sequence implements a $\pi$-pulse on the  $\ket{S}$ to $\ket{S'}$ transition.
Any electron population left in the D-state manifolds after spin echo due to e.g., imperfections in the $\pi$-pulses, is emptied in the beginning of the next photon generation attempt by the \SI{854}{nm} and \SI{866}{nm} light as described in the Loop 2 sequence.

Section \ref{sec:spin_echo_model} presents a model of the spin-echo performance.

 \subsubsection{Details on the DBSM of the ion-qubits} 
\label{DBSM}
In summary, the two ion-qubits are effectively projected into a basis of maximally entangled states by first applying an entangling `Molmer-Sorensen' (MS) gate followed by projectively measuring the individual ion-qubits in the logical basis, using the electron-shelving technique. 
The effective measurement is described by the set of four measurement operators that project into an orthogonal set of maximally entangled states $\{U^\dagger \ket{v}\bra{v} U\}$, where $\ket{v}$ are the four logical basis states of two qubits and $U=\exp(-i\sigma_\zeta^1\sigma_\zeta^2 \pi/4)$. Here, $\sigma_\zeta^k = \cos(\zeta)\sigma_x^k+\sin(\zeta)\sigma_y^k$, where $\sigma_{x(y)}^k$ is the $x(y)$ Pauli operator on qubit $k$ and $\zeta$ is defined by the phases of the laser fields driving the MS gate. For example, obtaining the final ion logical basis outcome $\ket{00}$  corresponds to an effective projection of  the state of the ions before the MS operation into the  maximally entangled state given by $U^\dagger \ket{00}= 1/\sqrt{2}(\ket{00}+i\ket{11})$ (assuming $\zeta = 0$). 

The MS gate is implemented on two ion-qubits, where each qubit is encoded into the $\ket{S}/\ket{D}$ states.
Specifically, the gate is realised with a 107$~\mu$s-long laser pulse at \SI{729}{nm} that contains three frequency components (tones). The first tone is $\delta=8.7$ kHz below the first red axial center-of-mass sideband of the $\ket{S}$ to $\ket{D}$ transition. The second is $\delta$  above the first blue axial center-of-mass sideband of the $\ket{S}$ to $\ket{D}$ transition. The third field is blue detuned from the $\ket{S}$ to $\ket{D}$ transition by 138 kHz. 

The two `symmetrically detuned' tones off-resonantly drive the axial COM mode of the ions in a way that depends on the qubit state, and, with appropriately set Rabi frequencies, yield a maximally-entangling qubit-qubit gate at a time of $1/\delta$: the time at which the motional state ideally returns to its initial value.  Those tones alone cause an AC Stark shift of the $\ket{S}$ to $\ket{D}$ transition (predominantly due to far off-resonant dipole transitions \cite{Kirchmairthesis}). The function of the third field is to compensate for that AC Stark shift by generating an equal and opposite AC Stark shift directly on the $\ket{S}$ to $\ket{D}$ transition. 

The MS gate quality is partially characterised by applying it to two ion-qubits in the state $\ket{00}=\ket{S,S}$ and determining the fidelity with which a maximally entangled state of the form {$\ket{\psi(\zeta)}=1/\sqrt{2}(\ket{00}-ie^{2i\zeta}\ket{11})$} is generated.
In the lab, the two qubit state $\rho$ is generated. The fidelity  $F = Tr(\ket{\psi}\bra{\psi}\rho)$ is determined by measuring the logical state probabilities and the amplitude of the parity oscillations, as described, for example, in \cite{Kirchmairthesis}. We characterise the gate in two cases: when the axial COM mode is ground state cooled, which corresponds to the conditions during Loop 1 of the repeater protocol, and;  when the axial COM mode is Doppler cooled which corresponds approximately to the conditions expected at the end of Loop 2 in the repeater protocol.  In both cases, the MS gate (with the same parameters) produces a state with fidelity of F = 0.95(2). Here, the uncertainty accounts for the gate performance variation between the gate calibrations.   
The phase of the state {$\zeta$}  for the two cases above was measured to differ by no more than $5^\circ$.

The ion-string is heated by photon scattering during the photon generation attempts.  Therefore, robustness of the quantum logic operations, including the MS gate, to different ion string temperatures is critical in our current repeater implementation.
See sec.  \ref{sec:temp} for more details on the motional state of the ions during different stages of the repeater protocol. 
For a discussion of various methods to compensate the AC Stark shifts in MS gates see section 3.5.1. of \cite{Kirchmairthesis}.

Readout of the logical state of the ions is performed via the electron shelving technique, during which both ions are illuminate by laser fields at \SI{397}{nm} and \SI{866}{nm}. Atomic fluorescence at 397nm from the individual ions is imaged onto a digital camera at \SI{397}{nm}, where the individual ions are well resolved. A bright fluorescing ion signals a qubit in the state $\ket{0}=\ket{S}$ while a dark, non-fluorescing ion signals the qubit state $\ket{1}=\ket{D}$.  Two milliseconds of electron shelving is performed per measurement. \\

\section{Main results}

\subsection{Tomographic reconstruction of two-photon states}
\label{phot_phot_reconstruction}

A total of eight two-photon density matrices are reconstructed, which are produced by the repeater protocol. Four states correspond to the case where ion A stored a qubit in memory, $\rho_{A,i}$, and four states correspond to the cases when ion B stored a qubit in memory, $\rho_{B,i}$, where $i \in$[1,2,3,4] denotes the ion DBSM outcome. Details on how the states states $\rho_{A,i}$ were reconstructed  are now given. The states $\rho_{B,i}$ were reconstructed in analogous fashion. 

Tomographic reconstruction of two-photon polarization states was performed by projecting each photon at each node into one of the basis states of three polarization bases: horizontal/vertical, diagonal/anti-diagonal, and left/right circular  (H/V, D/A, and R/L, respectively). 
For the two polarization qubits there are nine possible combined measurement bases. 
The 36 possible outcomes are described by a corresponding set of 36 projective operators $\{ O_j \}$. 
At each node there is one photon detector (Figure 2). We first set the waveplates to the angles corresponding to an HH projection and run the repeater protocol $N$ times, then we switch to the next projection and run the repeater protocol $N$ times and repeat for all 36 projections. 
For tomographic reconstruction, experimental estimates of the probabilities for obtaining these 36 outcomes conditioned on the ion outcome $i$ are required.

Consider the four outcomes associated with measuring the photons at each node in the H/V basis. 
Those probabilities for obtaining those outcomes are $p(HH), p(HV), p(VH), p(VV)$ where the first letter in brackets stands for the detected polarization state at the Node A and the second at the Node B, and $p(HH)+ p(HV)+ p(VH)+ p(VV) = 1$. 
If both H and V outcomes could be detected at the nodes with the same efficiency (e.g., using two identical detectors at each Node, with one in each output port of the polarising beam splitter) then those probabilities could be determined as follows. For example, $p(HH) = C_{i,HH}/(C_{i,HH}+C_{i,HV}+C_{i,VH}+C_{i,VV})$ where each term is the number of two photon coincident detection events of the labeled polarization, given the ion measurement outcome $i$.
All the other probabilities discussed could be obtained analogously. Since we do not have two detectors at each node, the aforementioned approach to obtaining probabilities is not possible.

In our experiment only photon detector is present at each node. Therefore, the probability to detect a photon per attempt at either node depends on the photon polarization state and the projection chosen. Moreover, the numbers of attempts executed in Loop 1 and Loop 2 during a single run of the repeater protocol depend on the probability to detect the photons at the nodes. 
As a result of last two statements, the total number of attempts executed in $N$ runs of the repeater protocol differs for each of the projections. For example, more attempts to generate photons will be made in the case of a lower detection probability. 
To recover the probabilities discussed it is therefore not enough to know only the numbers of coincidence detection events: the number of attempts also has to be known for each projection measurement. 
However, we do not determine the number of attempts (including the failed ones) that lead specifically to the ion state outcome $i$ as our protocol does not perform measurement of the ions states (DBSM) when no photon was detected. Performing that measurement after every attempt would be extremely time consuming.

We tackle the aforementioned problem with a Bayesian approach. 
Let $P(HH|i)$ be the desired conditional probability to find two photons in the state HH given that the ions are in the state $i$.  
According to the Bayes theorem  $P(HH|i) = P(i|HH)\times P(HH)/P(i)$ where $ P(i|HH)$ is the probability of the ions outcome  $i$ when the photons are detected at both  nodes in the polarization state H, $P(HH)$ is the probability to find the photons in the state HH regardless the state of the ions and $P(i)$ is the probability of the ions outcome $i$. 
From the data we calculate $P(i|HH) = C_{i,HH}/(C_{1,HH}+C_{2,HH}+C_{3,HH}+C_{4,HH})$ where $C_{i,HH}$ is the number of two-photon success events that lead to the ions outcome $i$ and  were recorded for the polarization analysis projection HH.
To determine   $P(HH)$ we again use the conditional probabilities, i.e. $P(HH) =P({A,H})\times P(B,H|A,H) $ where $P({A,H})$ is the probability to detect a single photon at the node A in the chosen polarization projection (H) and $P({B,H}|A,H)$ is the conditional probability to detect the photon at node B during loop 2 in the chosen projection given that the photon at node A is already detected.
From the experimental data we calculate  $P({A,H}) = N_{A,H}/M_{A,H}$ where $N_{A,H}$ is the total number of detection events at node A during loop 1  while the polarization analysis at that node was set to $H$ projection and  $M_{A,H}$ is the total number of Raman laser pulses addressed to the  ion A while the polarization analysis at that node was set to $H$.  
$P({B,H}|A,H) = C_{HH}/M_{B,H|A,H}$ where $C_{HH}$ is the number of two-photon events for the polarization setting HH and any ions state; $M_{B,H|A,H}$ is the number of Raman pulses on the ion B during loop 2 following successful detection of a photon at node A with the polarization H  during loop 1.
Combining what was introduced in this paragraph one obtains  $P(HH|i) = C_{i,HH}/(C_{1,HH}+C_{2,HH}+C_{3,HH}+C_{4,HH})\times N_{A,H}/M_{A,H}\times C_{HH}/M_{B,H|A,H}/P(i)$ which can be simplified 
to  
\begin{equation}
P(HH|i) = C_{i,HH}\times N_{A,H}/M_{A,H}/M_{B,H|A,H}/ P(i)
\label{eq_probs}
\end{equation}

Following the procedure of the previous paragraph we obtain 2-photon detection probabilities for all 4 projections  for a given basis and the given  outcome of the ions state detection, i.e. $P(HH|i), P(HV|i), P(VH|i), P(VV|i)$. 
We normalize the obtained probabilities values to satisfy $P(HH|i)+P(HV|i)+P(VH|i)+P(VV|i) = 1$, the common factor $P(i)$ from the eq. (\ref{eq_probs}) drops out at this step.
The probabilities for all 9 measurement bases obtained in the way described above are then used to reconstruct the two-qubit density matrix by  maximum-likelihood method. 
Specifically, following \cite{PRA_2003_Jezek_likelihood}, we numerically search for the two-qubit state (density matrix $\rho$) that for the set of projectors $\{ O_j \}$ would most likely provide the outcomes that were observed in the experiment.

\subsection{Locally rotating the two-photon density matrices into the Bell state form}
\label{localrotations}

Section \ref{phot_phot_reconstruction} explains how the two-photon states $\rho_{A,i}$ and $\rho_{B,i}$ are reconstructed from the raw data. As a reminder: the index $i$ stands for the obtained ion-measurement outcome (taking a value from 1 to 4) and the letter (A or B) specifies which ion stored a qubit in memory. 

Figure 3c in the main text presents four two-photon density matrices which we will now label as $\rho^{Bell}_{B,i}$. Those matrices have been rotated, from the ones directly reconstructed from the raw data, by the same two single-qubit rotations. That is,  $\rho^{Bell}_{B,i}=(U_B)\rho_{B,i}(U_{B})^{\dagger}$, where $U_B = u_{B1}\otimes u_{B2}$ and  $u_{B1}$ and $u_{B2}$ are single-qubit unitaries on photon-qubit 1 and 2, respectively.  Therefore, $U_B$ is a local unitary operator and does not change the entanglement content of the states $\rho_{B,i}$. Moreover, since $U_B$ does not depend on $i$, the inner product of the states $\rho_{B,i}$ and of the states $\rho^{Bell}_{B,i}$ is unchanged.  In analogy, we also define rotated two-photon density matrices $\rho^{Bell}_{A,i}$, where $\rho^{Bell}_{A,i}=(U_A)\rho_{A,i}(U_{A})^{\dagger}$ and $U_A = u_{A1}\otimes u_{A2}$. 

Next, we explain how the local rotations $U_A $ and $U_B$ are obtained. The method to obtain $U_B$ will be given as an example, and the method to obtain $U_A$ is done analogously. 
First, we define the fidelity $F_i= Tr(\ket{B_i}\bra{B_i}(U_B)\rho_{B,i}(U_{B})^{\dagger})$, where $\ket{B_i}$ are the four Bell states. 
Specifically $\ket{B_1} = \Phi^+ = 1/\sqrt{2}(\ket{00}+\ket{11})$, $\ket{B_2} = \Phi^- = 1/\sqrt{2}(\ket{00}-\ket{11})$, $\ket{B_3} = \Psi^+ = 1/\sqrt{2}(\ket{10}+\ket{01})$, $\ket{B_4} = \Psi^- = 1/\sqrt{2}(\ket{01}-\ket{10})$ where $\ket{0}(\ket{1})$ stands for the horizontal (vertical) linear polarization state.

Finally, the function $4-\sum_i F_i^2$ is minimised by numerical search, where the minimisation is over the parameters of the arbitrary local rotation $U_{B}$.

\subsection{Numerical simulation of feed forward}
\label{sec:feedforward}

The repeater delivers eight different entangled two-photon states $\rho_{A(B),i}$ to the photonic nodes (\ref{phot_phot_reconstruction}). Which state is delivered is heralded by which of the four ion outcomes is obtained ($i$) and which ion stored a qubit in memory (A or B). 
Those eight states could be converted into a single state by implementing outcome-dependent single-qubit rotations of the distributed two-qubit states: a process known as feedforward. In the following, we explain how our numerical simulation of that feedforward was done, yielding a single state delivered by the protocol with a fidelity of $72^{+2}_{-2}\%$ with respect to the Bell state $\ket{\Phi ^+ } = 1/\sqrt(2)(\ket{00}+\ket{11})$. 
Our approach is to apply the outcome-dependant single qubit rotations to the operators used in the tomographic state reconstruction, which allows a single state to be obtained from a single maximum likelihood process. 
In the future, the entanglement distributed by the repeater node should be stored in remote quantum matter (e.g., other repeater nodes), in which case the feedforward can be implemented experimentally.

Recall the set of 36 projectors $\{O_j\}$ introduced in \ref{phot_phot_reconstruction}, which are used to reconstruct each of the eight two-photon states $\rho_{A(B),i}$. We define 8 new sets of 36 projectors  $\{ O_{A(B),i}\} =  \{ (S_{i}U_{A(B)})^\dagger O  (U_{A(B)}S_{i})\}$, one set for each data set that is used to reconstruct $\rho_{A(B),i}$.
Here, $U_{A(B)}$ are the local rotations that convert the raw reconstructed states $\rho_{A(B),i}$ into the Bell-form, as described in \refsec{localrotations}. 
Furthermore, $S_i \in [I\otimes I, I \otimes \sigma_z, I \otimes \sigma_x, I \otimes \sigma_y]$, where $I$ is the one qubit identity matrix and $\sigma_{j}$ are the Pauli matrices. The $S_i$ operators convert all the Bell states $\ket{B_i}$ into the state $\ket{\Phi ^+}$. 
Next, a single tomographic reconstruction is performed via the maximum likelihood technique over a concatenated set of all $8\times 36$ projective measurements operators and correspondingly concatenated data sets, yielding a single state $\rho_{ff}$. The fidelity of $72^{+2}_{-2}\%$, reported in the main text, is calculated as $F_{ff}=Tr(\ket{\Psi ^+}\bra{\Psi ^+}\rho_{ff})$.

The described feedforward simulation was repeated, but this time using the first half of the complete data set to find the local unitary operators $U_{A(B)}$ and then the second half of the complete data to reconstruct the state $\rho_{ff}$. The result is $F_{ff}=71^{+4}_{-3}\%$ proving that the entangled state can be distributed with independently-calibrated local rotations.

\subsection{Quantities derived from the two-photon states} 
\label{sec:probs_and_MC}

The Monte-Carlo method is used to determine the statistical uncertainties in quantities derived from tomographically-reconstructed density matrices.   
Section \ref{phot_phot_reconstruction} describes how each two-photon state is reconstructed from a data set containing  vectors of two-photon detection events and single-photon detection events  together with the numbers of attempts.
For each data set that is used to reconstruct a density matrix, we numerically generate $Mc = 500$ vectors of `noisy' two-photon detection events and $Mc = 500$ vectors of `noisy'  single photon detection events. 
%
%
{The $n^{th}$ element of each of the 500 noisy vectors is drawn randomly from a simulated Poissonian distribution, with mean value equal to the experimentally recorded value for that element.}
Whenever the experimentally recorded number of detection event is zero, that value is substituted with a one to allow for statistical fluctuations between the `noisy' sets also in these cases and avoid underestimation of the errors. 
For example, for the data presented in the Figure 3 of the main text, in which two-photon detection events were recorded for 36 polarisation projectors and 4 different ion outcomes, only two times was zero two-photon counts recorded.
From these noisy data sets we derive outcome probabilities, the same way as we do for the experimentally obtained data set. 
Next we reconstruct the $Mc$ density matrices for this noisy data via a maximum liklihood search.
The result is that for any state reconstructed directly from the raw data, $\rho_{exp}$,  we have $Mc$ reconstructed noisy states. 

For each state $\rho_{exp}$ the value of some quantity of interest $c$ is calculated: these are the values given in the main text and the supplementary material. To derive error bars, we first calculate the quantity of interest for the associated $Mc$  noisy states, yielding a distribution $D$ of values with median value $c_0$. 
We then numerically find values $\delta_-$ and $\delta+$ such that $c_0-\delta_-$ and  $c_0+\delta_+$ represent the  same quantiles of the distribution $D(c)$ as  the mean minus standard deviation and the mean plus standard deviation represent for the normal distribution,  respectively.  
That is, {$0.1590$} of the  distribution $D$ lies above $c_0-\delta_-$ ( below $c_0+\delta_+$). 
Error bars are given as $c\pm ^{\delta_+}_{\delta_-} $ {which, in the case of a normal distribution, corresponds to $\pm$1 standard deviation of uncertainty.} 

As an example, consider the fidelities presented in the caption Figure 3c of the main text: fidelities between the  two-photon states where ion B stores a qubit in memory and target Bell states,  $F_{B,i}^{Bell}=Tr(\ket{B_i}\bra{B_i} \rho^{Bell}_{B,i})$. 
Figure \ref{figsup:fid_sep2} presents the four values $F_{B,i}^{Bell}$ (vertical green lines) on top of the histogram of equivalent values calculated for the $Mc$ noisy data sets. 
Figure \ref{figsup:fid_sep1} presents the same information, but calculated  for the state where ion A stored a memory qubit ($\rho_{A,i}$).  {Note, that the rotations $U_B$ ($U_A$), which are used to find  $\rho^{Bell}_{B,i}$ ($\rho^{Bell}_{A,i}$) according to \refsec{localrotations}, are found only once: For the experimental data set, without resampling.} 

As a second example, histograms of the concurrences of the states $\rho_{B,i}$  and $\rho_{A,i}$ are provided in Figures \ref{figsup:conc_sep2} ans  \ref{figsup:conc_sep1} respectively. The resultant concurrences in those figures are   $[50^{+14}_{-12} ,43^{+14}_{-15},44^{+13}_{-12}, 72^{+10}_{-11}]\%$ ($[27^{+14}_{-15} ,43^{+12}_{-13},28^{+14}_{-13}, 66^{+12}_{-12}]\%$).
The concurrence is a measure of entanglement for two-qubit density matrices that is invariant under local transformations. A concurrence value greater than zero proves entanglement and the maximum value is 1 \cite{Hill97_concurrence}.

\subsection{Efficiencies and rates in the repeater protocol}
\subsubsection{Efficiency budget for single photon detection at the end nodes}
\label{sec:efficiency}

The probability to detect a photon at Node {A} during Loop 1 of the repeater protocol is $P_{A,1} = 3.06(5)\times 10^{-3}$.
The probability to detect a photon at Node B during Loop 1 of the repeater protocol is $P_{B,1} = 2.36(4)\times 10^{-3}$.
The corresponding probabilities for Loop 2 are $P_{A,2} = 2.64(8)\times 10^{-3}$ ($P_{B,2} = 1.81(6)\times 10^{-3})$.
Two causes of the lower photon detection probabilities during the Loop 2 are now described. However, we stress that the drop in photon detection efficiency per attempt in Loop 2 is remarkable small (Figure 3b, main text), even after 190 attempts. 

First, since optical pumping is not performed in Loop 2, the ion in the focus of the Raman laser is not reinitialised in the state $\ket{S}$ after each photon generation attempt. As such, at the beginning of each photon generation attempt, that ion is in a partially mixed state across the $\ket{S/S'}$ manifold. 
However, the addressed Raman laser can cause population transfer from $\ket{S'}$ to $\ket{S}$ via 
spontaneous scattering from various states within the $\ket{4^{2} P_{3/2}}$ manifold. Therefore, a cavity photon can still be generated in cases where the ion starts in $\ket{S'}$. After each Raman pulse in the loop 2, the probability of the ion state remaining in the $\ket{S'}$ to $\ket{S}$ manifold is vanishingly small. Since the $\ket{P}$ state can also (weakly) decay to the $\ket{D_{5/2}}$ and $\ket{D_{3/2}}$   manifolds, scattering processes also slightly reduce the overall photon generation probability. 

Second, the increasing temperature of the ion string during Loop 2, due to photon absorption and scattering. The most significant affect of this heating is to reduce the coupling of the ultraviolet-wavelength Raman laser to the desired atomic transition in the photon generation process. Consequently the coupling strength of the BCMRT photon generation process is reduces, while the rate of spontaneous emission remains the same, yielding a reduced photon generation probability. 
Measurements and simulations of the temperature change of the ion string during Loop 2 is presented in sec. \refsec{sec:temp}.

A detailed efficiency budget list for single photon detection during Loop 1 is now given. When not given explicitly, the uncertainty in quote probabilities are half of the last significant digit. The beginning of each element in the following list gives the probability associated with a distinct part or process in the experiment, beginning with photon generation and ending in photon detection.

\begin{enumerate}
\item 0.52: probability for photon generation into the cavity mode within the relevant 40 $\mu s$ window. A value of 0.61 is obtained from numerical simulations with independently measured parameters, see sec. \ref{sec:WP_simulation}. However, comparison between the single photon wavepackets predicted by those simulations, and the ones detected in the experiment (Figure 3a, main text), suggest that the ion-cavity coupling strength in the repeater protocol was lower than anticipated. 
Separately, we note that this probability, for photon emission into the cavity, is far lower than has been proven possible in our system (0.940(2) \cite{SchuppPRX2021}). 
We expect that the repeater value of 0.52 could be significantly improved upon in future by changes to the setup including rotating the principle magnetic field axis and Raman laser polarization to optimise the relevant laser-atom coupling strength. 
\item 0.78(2): probability that, once a cavity photon is emitted into the cavity, the cavity photon exist the cavity into freespace on the other side of the output mirror \cite{SchuppPRX2021}.   
\item 0.96(1): transmission of free-space optical elements between cavity output mirror and first fiber coupler, see $P_{el}$ in \cite{SchuppPRX2021}.
\item 0.81(3): efficiency of coupling the photons to the first single mode fiber \cite{SchuppPRX2021}. This value should rather be considered an upper bound, since it was measured some days before the data presented in this paper was taken. We anticipate that coupling could be  improved in future with better couplers and anti-reflection coated fiber end facet. 
\item 0.23(1): fiber-coupled input to fiber-coupled output efficiency of the telecom conversion setup. This value was measured with classical light a few hours before executing the repeater protocol. The maximum observed value was 0.27.  
\item 0.55(0.5) and 0.46(0.5): efficiencies of the photon paths from the output of the conversion setup to the fibers connected to   single photon detectors (SPDs) for Nodes A and B, respectively. The paths include the photon switch (0.86 and 0.82 transmission efficiency, respectively), out-coupling to the free-space polarization analysis elements and coupling to the fibers of SPDs. The paths here do not include the long fiber spools. 
\item 0.36(0.5) and 0.42(0.5): transmissions of 25 km SMF-28 fiber spools including one fiber connector each, for Nodes A and B, respectively.  as measured with classical light.
\item  0.5: theoretical probability of photon transmission through the polarising beam splitter in each photonic node, for a maximally entangled ion-photon state. 
\item  0.71 and 0.56: combined efficiencies of the 2 and 4 fiber joiners in the paths to nodes A and B, respectively that were not accounted elsewhere in this list. This efficiency is the lower bound according to the datasheets provided by the manufacturer. \item  0.75(2): detection efficiency of either of the telecom single photon detectors.
\end{enumerate}

The product of the above listed efficiencies in Loop 1 are $3.8(3)\times 10^{-3}$ and $2.9(2)\times 10^{-3}$ for Nodes A and B, respectively. The given error is propagated as for independent random variables. 
These calibrated values agree with the corresponding measured ones, given at the start of this section. The discrepancy beyond one standard deviation can be a result of limited stability of the values provided in the list during the time between the calibration and the repeater protocol execution.

\subsubsection{$P^0_{link}$: Photon detection efficiency without fiber loss}
\label{0km}

Consider the total probability $P^0_{link}$ to generate, deliver and detect a telecom photon at a photonic end node, excluding only the transmission losses of the total length of fiber through which the photon travel between neighbouring nodes.  $P^0_{link}$ is a key parameter for characterising and modelling the performance of the ion node. 

For photonic nodes A and B one could determine $P^0_{link}$ values from the measured single-photon efficiencies in each loop separately (e.g., $P_{A,1}$ and $P_{A,2}$) and divide by the the measured transmission probabilities of the 25km fiber spools (sec. \ref{sec:efficiency}). However, to simplify the analysis of \refsec{sec:SKR}, we do not extract the probabilities separately for different loops and neglect asymmetry between the different photonic nodes. We thus define $P^0_{link}$ as the total number of photons recorded during the execution of the repeater protocol divided by the total number of the Raman pulses and by the average transmission of the two \SI{25}{km} fiber spools, yielding $P^0_{link} = 0.0065(6)$.

\subsection{Repeater versus direct-transmission configuration: rate comparison}
\label{sec:compare_direct}

\subsubsection{Experimental details and results}

In the main text we describe an experimental comparison of performance when the ion-node is used in repeater configuration with performance when the ion-node is used in a direct transmission configuration. In the latter case, the ion-node (everything in the box labeled 'repeater node' in Figure 2) is used as a source of ion-photon entanglement and placed at one end of the joined \SI{25}{km} fibers (total \SI{50}{km}) with photon Node A  at the other end. \\

\noindent \textbf {Rates}

\noindent As described in the main text, we ran the ion-node in a direct transmission configuration and recorded the success rate for the establishment of ion-photon entanglement over 50 km, without photon polarization analysis. Here, polarisation analysis is removed by connecting the final spool's output fiber to the detector in Node A. 
Afterwards, we ran the ion-node in repeater configuration and recorded the success rate for establishing photon-photon entanglement over 50 km, without photon polarization analysis. Here, polarisation analysis is removed by connecting the output fiber of each spool directly to the fiber-coupled detector in its corresponding node.

In repeater configuration, 2839 success were recorded over 8 minutes, yielding a raw success rate of 5.9 Hz. 
In the direct transmission configuration, 1569 successes were recorded over 6.8 minutes, yielding a raw success rate of 3.8 Hz. 
Next we consider only the `active times' during those experiments. The   `active times'  correspond to those in which either Loop 1, Loop 2 or ion state detection were being executed (the latter two are only relevant for the repeater configuration). Other times periods e.g., state initialization and  software compilation are excluded. 
When considering only active time, the achieved success rates are 9.2 and 6.7 Hz, for the repeater and direct configurations respectively. 
Note that during the active time, entanglement distribution attempts were performed with the rates of 2.5  and 1.3 kHz for the repeater and direct schemes, respectively, which are close to saturating the maximum possible attempts rates of 4 and 2 kHz set by the photon travel  and return signal times. \\

\noindent \textbf {Efficiencies}

\noindent Removal of the polarization analysis optics improves the overall efficiency of the repeater protocol configuration, compared to the results presented in Figure 3 of the main text. The polarization analysis optics is only required to characterise the delivered entangled states and costs 0.5 efficiency in each instance (since we do not have a photon detector at each output port and each photon is near maximally entangled to its ion). 
Given this improved efficiency, we now repeat some of the key efficiency calculations for the node. 

Photon detection efficiencies for the repeater protocol configuration without polarization optics are now given. First, following the method in \refsec{0km}, we measured $P^0_{link} = 0.018(1)$. 
To recap, the value of $P^0_{link}$ given here is the average probability to generate, deliver and detect a telecom photon at a photonic end node, excluding only the transmission losses of the \SI{25}{km} fiber between neighbouring nodes. This value is used as $P^0_{link}$ in all subsequent calculations involving this parameters in this supplementary material (see \refsec{sec:future} Future Systems).  Second, the total probability to detect a photon in Loop 2 is 0.68(1), which compares to compared to the 0.346(4) for the experiment with polarization analysis reported in the main text.  

For the direct transmission configuration without polarization optics we obtain a value of $P^{0}_{link} = 0.017(1)$, where there is only one photonic node involved. That value agrees with the value above for the repeater configuration. \\

\subsubsection{Numerical model for our repeater configuration rate advantage}
\label{sec:numeric_rates}

\noindent The achieved success rates when considering active time are 9.2 and 6.7 Hz for the repeater and direct configurations, respectively. The repeater configuration thus offers a rate advantage.  

We now describe a numerical model for predicting the success rates of our repeater and direct configurations over arbitrary channel length, and compare those predictions to the measured results. The model is set in the context of the application of secret key rate distribution. In that context, the success rates in our experiments are equivalent to a raw key rate (RKR). The meaning of a RKR is discussed later in \refsec{sec:SKR}.
  
The model for calculating the RKR contains the following parameters: $P^{0}_{link}$, as defined above; $L$, the total length of the fiber between the end nodes; $\gamma$, the attenuation of the fiber per unit length; $T_0$, the time for photon generation and reinitialisation in each attempt to distribute entanglement; K, the maximum number of photon generation attempts with a qubit in memory (Loop 2 attempts) and; $T_{swap}$, the time for the DBSM and feedforward. The last two parameters are only relevant for the repeater configuration. 

The modelled RKR is calculated as follows. 
First, we calculate the inverse of the (dimensionless) entanglement creation time which is defined and calculated in \cite{Sangouard2011_review} (See Appendix A sec.1,2).  Here, to account for the finite number of the Loop 2 (memory) attempts, we truncate the relevant infinite sum in \cite{Sangouard2011_review} at $n=K$. Moreover, we define the probability for successful qubit delivery in one attempt to an end node over fiber distance $L$ as\footnote{The quantity $P_s$ defined in Equation \ref{EqPs} is called $P_0$ in \cite{Sangouard2011_review}}

\begin{equation}
P_s = P_{link}^0\times 10^{-\gamma L}.
\label{EqPs}
\end{equation}

Second, we divide the inverse of the truncated entanglement creation time by the attempt duration time $T$ to obtain the RKR in Hz. $T$ is bounded by the light speed in the fiber $c$ as $T = T_0+L/c+T^{swap}$ for the repeater configuration and $T = T_0+2L/c$ for the direct configuration. $T^{swap}$ has been is included in the equation for $T$ for the repeater configuration since it is needed for the DBSM and feed forward.

The values of the model parameters used to produce the predicted RKR results in Figure \ref{fig:rates} are now given. We use: $P^0_{link} = 1.8\%$; $\gamma = 0.0173$ km$^{-1}$, from the average measured in the fiber spools used;  $T_0 = 123~\mu s$ and ($T_0 = 201~\mu s$) for the repeater and direct transmission configurations, respectively; $T^{swap} = 2157 \mu$s 
and; $K = 190$  as determined from the experiment. 
The $T_0$ value given above for the repeater configuration is determined from the sequence executed in Loop 2, where only one photon generation pulse is performed. The $T_0$ value given above for the direct transmission configuration is determined from the sequence executed in Loop 1. The time $T^{swap}$ should include both the time needed for the DBSM and feed-forward. In the model, the value of $T^{swap} = 2157 \mu$s is fixed, which is possible up to $L = 400$ km provided a fast enough DBSM (which should be entirely possible as no significant effort was made to reduce the time for electron shelving in the DBSM). 
For fiber lengths above 400 km, which are not presented in Figure \ref{fig:rates}, the photon travel time exceeds $2000 \mu$s, certainly requiring a bigger $T^{swap}$ value.
The value of $2157 \mu$s taken here corresponds to the current repeater protocol realization where no feed-forward was implemented. That value thus does not include at least $125~\mu$s needed for the feed-forward over 25 km of the fiber.

We now discuss the results of the numerical model. Figure \ref{fig:rates} presents the measured and predicted RKRs as a function of the telecom fiber channel length. 
Points with error bars show the two experimental RKRs (9.2 and 6.7 Hz for $L=$\SI{50}{km}) and dashed lines show the RKR from the numerical model. 
Note, the solid lines in Figure \ref{fig:rates} show secret key rates (SKR)s, calculated using a different model as described later in \refsec{sec:SKR} and should be ignored for now.

The RKR predictions of the model (dashed lines, Figure \ref{fig:rates}) are consistent with the data for $L =$ \SI{50}{km}. The repeater configuration is predicted to achieve a higher SKR than the direct configuration for fiber lengths above 30 km.  At 0 km, the repeater configuration is predicted to have lower rates than direct transmission since e.g., the repeater requires two single photon detection events for success compared to one for direct transmission.
For $L< 100$ km the dashed line for repeater configuration has a bigger slope than the one for the direct configuration. 
In other words, in this range the repeater configuration provides better RKR scaling with $L$ than the direct configuration.
However, due to the finite value used for K---the number of entanglement distribution attempts in Loop 2--- the slope of the RKR predicted for the repeater configuration tends to that for the direct configuration for increasing $L$. 
In order to maintain the advantageous slope (scaling) of the RKR for the repeater configuration for higher values of $L$, one can increase the value of $K$ used in the model.

\subsubsection{Analytical expressions for a repeater configuration rate advantage}

In the main text and previous subsection we discuss an experimental comparison of performance of the 50 km fiber channel with the ion repeater node in the middle and performance of the 50 km fiber channel where the ion node is placed at one end sending photons directly to another end. In the main text it is states that ``Analytic expressions for the performance requirements for a repeater node to beat itself in direct transmission are summarised in [Supp Mat]. Those expressions are now provided, following the approach of \cite{Sangouard2011_review} (see Appendix A of that work).

The rates predicted by the numerical model in the previous subsection can be expressed analytically in the case  $K \rightarrow \infty$ ($K$ is the maximum number of photon generation attempts with a qubit in memory). 
In the following we first summarize the derivation of those expressions. 
Then, using those expressions, we derive a bound on the average time, $\overline{t}$, that the repeater needs stores a qubit in memory for a protocol success.  
Finally, using parameters of our experimental system, the predictions of the expressions are presented. \\

\label{sec:analytic_rates}
 \noindent \textbf{Comparison of a node in repeater and direct configurations.}

 \noindent First consider the direct configuration. Specifically, a node containing two matter-qubits, which sends photons entangled with the matter-qubits, over a fiber with total length $L$ at photon propagation speed $c$. 
The attempt rate is taken to be $2\times c/(2L)$. Note that during each of these attempts, one photon, entangled with one of the nodes' qubits, is sent across, hence the first factor of two. 
This attempt rate corresponds to the inverse of the time needed for the photon to be delivered and the information about it to come back to the sending node, as was implemented in our experimental sequence by setting a sufficient delay. The photon generation time $T_0$ used in the numerical model of the previous section is thus set to 0 for simplicity. 
Then, the success rate for establishing remote matter-photon entanglement $R_d$ is: 
\begin{equation}
R_d = P_s\times 2c/(2L) = P^0_{link}\eta\times c/L,
\end{equation} 
where $P_s= P^0_{link}\eta$ is overall photon detection (success) probability in a single attempt, $P^0_{link}$ is the zero-distance success probability defined in \ref{0km}  and  $\eta \propto 10^{-\gamma L}$ is the channel transmission, i.e. the probability that a photon sent by the sending node will arrive to the receiving node over a fiber channel of the length $L$.

Now consider the repeater configuration, which has total fiber length $L$ between end nodes, the repeater node placed in the middle and sending one photon to each endnode. 
For the repeater configuration $P^{'}_s= P^0_{link}\sqrt{\eta}$, is the probability to deliver a photon to an end node from the repeater node. The fiber length of $L/2$ between the repeater and any of the end nodes defines the entanglement distribution attempts rate of $c/L$.
The success  rate $R_r$ for the repeater configuration in the approximation of unlimited memory and $P^{'}_s \ll 1$ is calculated in sec. 2 of appendix A of \cite{Sangouard2011_review} as
\begin{equation}
R_r = \frac{2P^{'}_s}{3}\times c/L = \frac{2P^0_{link}\sqrt{\eta}}{3}\times c/L
\label{eq:Rr}
\end{equation}

where the time needed for the DBSM and feed-forward has been neglected, which is justified for $P^{'}_s \ll 1$ as DBSM (and feed-forward) happens rarely in that case.  

The ratio of rates is
\begin{equation}
\frac{R_r}{R_d} = \frac{2P^0_{link}\sqrt{\eta}\times 2}{3P^0_{link}\eta\times 2} = \frac{2}{3\sqrt{\eta}}
\end{equation}
Therefore, the repeater configuration achieves higher rate than the direct configuration with the same node when the half-link transmission $\sqrt{\eta}<2/3$ (so that ${R_r}/{R_d}>1$). 
With $\eta = 10^{-\gamma L}$ the condition above gives the bound for the link length 
\begin{equation}
L > -2\log(2/3)/\gamma .
\label{eq:L}
\end{equation}
 One sees that the condition for a rate advantage, under the assumptions made in the model, is simply a matter of channel length and not a question of well the node performs.

Recall the quantity $\overline{t}$: the average time that the repeater node stores a qubit in memory for a protocol success. We now refer to  $\overline{t}$ as the qubit storage time. 
We can now calculate the bound on $\overline{t}$ required in order to achieve a repeater rate advantage ${R_r}/{R_d}>1$.
Given the success probability in an attempt $P^{'}_s$, the average number of attempts $\overline{k}$ in Loop 2 before success in the repeater protocol is $\overline{k}=1/P^{'}_s$. Using the definition of  $P^{'}_s$ and condition \ref{eq:L} for ${R_r}/{R_d}>1$ we obtain 

\begin{equation}
\overline{k}= 1/(P^0_{link}\sqrt{\eta})>3/(2P^0_{link}).
\label{khat}
\end{equation}

Performing $\overline{k}$ entanglement distribution attempts requires time of $\overline{t} = {\overline{k}L}/{c}$, yielding

\begin{equation}
\overline{t} >\frac{3\log(3/2)}{P^0_{link}c\gamma}.
\label{that}
\end{equation}

Where Eq. \ref{eq:L} has been employed. Equation \ref{that} gives the desired bound on $\overline{t}$ required in order to achieve a repeater rate advantage ${R_r}/{R_d}>1$.
One sees that $\overline{t}$ 
grows as inverse of the node's efficiency $P^0_{link}$.

Finally, using parameters of our experimental system, we calculate the bound on the total fiber length $L$ (Eq. \ref{eq:L}) and the average memory storage time $\overline{t}$ (Eq. \ref{that}) for a repeater advantage. Using $\gamma = 0.0173$ km$^{-1}$, the group velocity of light in the SMF-28 fiber and the $P^0_{link} = 0.018$ achieved in our setup (see \refsec{0km}) we find that $L>20$ km and $\overline{t}>10$ ms. Recall that in the experiment reported $L = 25$ km and $\tau = 62(3) \mathrm{ms} > \overline{t}$.
In conclusion, the predictions from the analytical bounds are consistent with our observation of a rate (RKR) advantage in repeater configuration. \\

 \noindent \textbf{Comparing a repeater node with a perfect direct link.}
\noindent We now estimate the condition under which a repeater-node could achieve a higher success rate for entanglement distribution than a `perfect' realisation of direct transmission, defined by a source of matter-photon entanglement that achieves $P^0_{link} = 1$. 
For such a perfect source in direct transmission, with two matter-based emitters that can be operated in parallel, the achievable success rate over a total distance $L$ would be:
$$R^{perfect}_d = P_s\times 2c/(2L) = \eta\times c/L$$
Using the  expression \ref{eq:Rr} for the imperfect repeater as before, we obtain the ratio of success rates $\frac{R_r}{R^{perfect}_d} = \frac{2P^0_{link}}{3\sqrt{\eta}}$. The condition for a repeater advantage (${R_r}/{R^{perfect}_d}>1$) is now
\begin{equation}
\sqrt{\eta}< 2P^0_{link}/3.
\label{eq:eta}
\end{equation}

We now estimate  $\overline{t}^{perfect}$:  the average time that the repeater node stores a qubit in memory for a protocol success under the condition  ${R_r}/{R^{perfect}_d}>1$.
As done above for $\overline{t}$, we write $\overline{t}^{perfect} = \overline{k}L/c$ and $\overline{k}=1/P_s$. Now by  using the inequality \ref{eq:eta} instead of \ref{eq:L} before and writing $P_s$ and $\eta$ explicitly  we obtain
\begin{equation}
\overline{t}^{perfect}>  -\frac{3\log(P^0_{link}\times 2/3)}{(P^0_{link})^2\times\gamma c}.
\label{eq:tmem_perfect}
\end{equation}
Equation \ref{eq:eta} gives the bound on $\overline{t}^{perfect}$ required in order to achieve a repeater rate advantage over a direct configuration with perfect nodes, ${R_r}/{R^{perfect}_d}>1$.
One sees that $\overline{t}^{perfect}$ 
grows as inverse of the squared node's efficiency $P^0_{link}$.

Finally, using parameters of our experimental system, we calculate the bound on the average memory storage time $\overline{t}^{perfect}$ (Eq. \ref{that}).
For the currently-achieved $P^0_{link} = 0.018$, eq. \ref{eq:tmem_perfect} yields $\tau > 5$~s, which is greater than the current coherence times of the memories in the ion node.
For $P^0_{link} = 0.21$, Eq. \ref{eq:tmem_perfect} yields $\overline{t}^{perfect}> 10 $ms.

\section{Supporting results}

\subsection{Tomographic reconstruction of ion-photon states}
\label{sec:ion_photon_procedure}

Tomographic reconstruction of two-qubit ion-photon states is performed following the method described in \cite{Krutyanskiy:2019cx}. In summary, experiments are repeated for a tomographically-complete set of polarization analysis settings that are described in Sec \ref{phot_phot_reconstruction} and for a complete set of ion-qubit measurements bases, corresponding to the eigenstates of the three Pauli matrices. The nearest physical state to the data is found via the maximum likelihood method. The characterised ion-qubit is always encoded in the $\ket{S}/\ket{D}$ states.  A 729 nm $\pi/2$ pulse, with adjustable phase, is used when measurements of the ion-qubit are to be carried out in the Pauli $x$ and $y$ bases.
The florescence state of individual ions are obtained using an EMCCD camera, in the same way as in the last step of the DBSM (\ref{sec:sequence}). 

Following the notation in the main paper, the tomographically reconstructed state of ion-qubit `A' and photon qubit `a' is labelled as $\rho_{Aa}$. The equivalent state for ion B and photon b is $\rho_{Bb}$. 
Since photons A and B travel through different optical fibers spools without a preset birefringence, states  $\rho_{Aa}$ and $\rho_{Bb}$ are not expected to be the same. \\

\subsubsection{Measured distributed ion-photon states}
\label{sec:repeater_ion-phot_states}

Measurements were made to tomographically reconstruct the ion-photon states $\rho_{Aa}$ and $\rho_{Bb}$ distributed over the full repeater protocol.  For those measurements, the repeater protocol is implemented with telecom conversion and fiber spools, but without the entangling MS gate.

Before tomographic reconstruction, the data are split into two groups. 
The first group contains those cases in which photon A was detected in Loop 1 and photon B was detected in Loop 2. This first group thus contains cases where only ion A stored its qubit in memory and, after tomographic reconstruction, yields two states  $\rho^{Amem}_{Aa}$ and $\rho^{Amem}_{Bb}$.
The second group contains those cases in which photon B was detected in Loop 1 and photon A was detected in Loop 2. This second group thus contains cases where only ion B stored its qubit in memory and, after tomographic reconstruction, yields two states  $\rho^{Bmem}_{Aa}$ and $\rho^{Bmem}_{Bb}$.

To quantify the quality of the four ion-photon states from above we take their fidelities with the nearest maximum entangled state, in analogy to the method for photon-photon states described in \refsec{localrotations}. 
For example, for the state $\rho^{Amem}_{Aa}$ we define the arbitrary bi-local unitary rotation $U^{Amem}_{Aa} = u^{Amem}_{A}\otimes u^{Amem}_{a}$, where  $u^{Amem}_{A}$ and $u^{Amem}_{a}$ are arbitrary single-qubit unitaries on the ion (A) and photon (a) qubits respectively. 
We define the fidelity $F^{Amem}_{Aa} = Tr\left(\ket{(U^{Amem}_{Aa})^\dagger\Phi^+}\bra{\Phi^+U^{Amem}_{Aa}}\rho^{Amem}_{Aa}\right)$, where $\ket{\Phi^+}$ is the Bell state.
We then find, by numerical search, the $U^{Amem, Max}_{Aa}$ that maximises $F^{Amem}_{Aa}$ over all possible rotations $U^{Amem}_{Aa}$.
The state $\ket{(U^{Amem}_{Aa})^\dagger\Phi^+}$ is the nearest maximum entangled state to the state $\rho^{Amem}_{Aa}$. Finally we calculate the fidelity $F^{max}$ between that state and $\rho^{Amem}_{Aa}$.
By repeating the operations described for all four states $[\rho^{Amem}_{Aa},\rho^{Amem}_{Bb} \rho^{Bmem}_{Aa},\rho^{Bmem}_{Bb} ]$ we obtain $F^{max} = [0.77(7), 0.94(6), 0.88(6), 0.93(7)]$, each associated with a bi-local two-qubit unitary rotation.

We now discuss the origin of the differences in the obtained bi-local two-qubit unitary rotations. First, as already stated, the photons (a and b) travel through different fibers each obtaining a different uncalibrated constant polarization rotation. 
Second, we expect that the ion qubit that was not used as a memory qubit in Loop 2 obtains an unknown Pauli Z rotation for the reasons discussed in the next paragraph. 
We repeat the above four numerical searches, but this time restrict the bi-local rotations to   
$U^{Amem, \prime}_{Aa} = I \otimes u_{a}$,
$U^{Amem, \prime}_{Bb} = e^{i\sigma_z\theta_{Amem}}\otimes u_{b}$,
$U^{Bmem, \prime}_{Aa} = e^{i\sigma_z\theta_{Bmem}} \otimes u_{a}$,
$U^{Bmem, \prime}_{Bb} = I\otimes u_{b}$, where $u_a(b)$ are the polarization rotations in the fibers to the nodes A(B) and $\theta_{Amem(Bmem)}$ is the angle of the ion qubit's Pauli Z rotations.
We then define fidelity $F^\prime$ analogously to $F^{max}$ 
and search for the rotations  
that minimize function $4-\sum{F^\prime}$.

We now quantify how well the restricted rotations match the ones that allow for arbitrary bi-local rotations. 
We calculate the vector of four fidelities between pure states:  
$[F(\ket{(U^{Amem}_{Aa})^\dagger\Phi^+},\ket{(U^{Amem, \prime}_{Aa})^\dagger\Phi^+}), 
F(\ket{(U^{Amem}_{Bb})^\dagger\Phi^+},$ $\ket{(U^{Amem, \prime}_{Bb})^\dagger\Phi^+}), 
etc]$, obtaining the values $[0.99(1), 0.99(1),0.99(1),0.99(1)]$, respectively. We therefore conclude that differences between the tomographically-reconstructed ion-photon states are well described by the rotations of the polarization in the fibers and rotations of the ion qubit generated by the Pauli $\sigma_z$ operator. We find, up to the $2\pi$ periodicity,   $\theta_{Amem} = 1.4(2)$ rad and $\theta_{Bmem} = 1.4(3)$ rad. The uncertainties given are calculated by the Monte-Carlo approach similar to \refsec{sec:probs_and_MC}. Specifically, here we calculate the quantities of interest for an array of noisy (ion-photon) data sets and use the standard deviation of the quantity of interest as the uncertainty. 

The expected origin of the aforementioned ion qubit Pauli $\sigma_z$ rotations is now described. During Loop 2 the ions are exposed to \SI{806}{nm} light used for locking the length of the cavity around the ions. That exposure occurs during the 125 $\mu$s waiting time after the photon generation laser pulse in each attempt of Loop 2. During that exposure, the ion on which photon generation attempts are made in Loop 2 has a qubit encoded in the $D/D^\prime$ states. Those states experience an AC Stark shift due to the \SI{806}{nm} light which we estimate can result in the aforementioned state rotations. Precise calculations are not possible as we do not have precise information about the location of the two ions with respect to the cavity standing wave at \SI{806}{nm}.

Straightforward calculations for the case in which the ion-qubits did not undergo the rotations characterised by angles $\theta_{Amem}$ and $\theta_{Bmem}$, that is $\theta_{Amem}=\theta_{Bmem}=0$, show that the repeater protocol should yield a total of four orthogonal two-photon states shared by nodes A and B. Those four states are obtained conditional on the logical ion-qubit measurement outcome. 
Furthermore, in the case in which $\theta_{Amem}=\theta_{Bmem}\neq 0$, and perfect ion-photon Bell states, we find that the repeater protocol should yield a total of 6 different two-photon states.  Those six states are obtained conditional on the logical ion-qubit measurement outcome and on which ion stored a qubit in memory. 
Finally, for the case in which  
$\theta_{Amem}\neq\theta_{Bmem}$, $\theta_{Amem}\neq 0$ and $\theta_{Bmem} \neq 0$, and perfect ion-photon Bell states, we find that the repeater protocol should yield a total of 8 different two-photon states. Those eight states are again obtained conditional on the logical ion-qubit measurement outcome and on which ion stored a qubit in memory. 
Although the measured values of $\theta_{Amem} = 1.4(2)$ rad and $\theta_{Bmem} = 1.4(3)$ rad are equal to within statistical uncertainty, they are not expected to be identical: the underlying Stark shifts expected to cause them have a strong spatial dependence and the ions were not positioned with respect to it. Therefore, we analysed our data by splitting it into 8 different sets of states.

\subsection{Characterisation of the ion-memories}
\label{sec:memory}
\subsubsection{Characterisation of the ion-photon states stored in memory}
\label{first}

To identify origins of protocol infidelity, the ion-photon entangled states stored in memory are characterised in a separate experiment. Here, a modified repeater protocol is performed: up to 30 Loop 1 attempts are made to detect a photon from ion A only. In successful cases, ion-qubit A is stored in memory, then a fixed number of Loop 2 attempts $k$ are made to generate a photon from ion B. Finally, the ion-photon state stored in memory, $\rho_{Aa}(k)$,  at each step $k$ of Loop 2 is tomographically reconstructed. 

This experiment is performed without fiber spools and telecom conversion for improved efficiencies, but wait times are included allowing for twice 25 km of photon travel. Consequently, the photons are detected at \SI{854}{nm} using an analogous optical setup to the telecom photonic nodes in Figure 2 of the main paper, involving a single superconducting nanowire detector in each node (manufactured by Scontel). The datasheet detection efficiencies for the two \SI{854}{nm} detectors are 87\% and 88\% with free running dark count rates of 0.3(1) Hz and 0.5(1) Hz. The fiber-coupled telecom switch is replaced by one suitable for \SI{854}{nm} (Photonwares). 

Figure 4 of the main text shows how the fidelity $F(\rho_{Aa}(k))$ of the measured $\rho_{Aa}(k)$ states—w.r.t. to the maximally-entangled state nearest to $\rho_{Aa}(0)$—evolve. The plotted fidelities values for different qubit storage times  are $\left\lbrace 0.96(2), 0.87(2), 0.74(2), 0.64(2)\right\rbrace $. These fidelities are, to within statistical uncertainty, limited by the purity ($Tr(\rho^2)$) of the states. That is, for any two-qubit state $\rho$, it is straightforward to prove that the maximum fidelity that $\rho$ can have with any maximally-entangled two-qubit state is less than or equal to $\sqrt{[Tr(\rho^2)]}$. For the reconstructed states, $\sqrt{[Tr(\rho_{Aa}(k)^2)]}= \left\lbrace 0.96(2), 0.88(2), 0.76(2), 0.69(2)\right\rbrace$, respectively. The observed fidelity drop with $k$ is thus dominated by non-unitary processes: processes that reduce the purity.

\subsubsection{Model of the dephasing dynamics of the ion-photon states stored in memory}

Figure 4 of the main paper contains a model of the fidelity dynamics of the ion-photon state stored in memory. 
The modelled fidelity, after memory storage time $t$, is a property of the modelled state $\tilde{\rho}_{Aa}(t, \tau)$, where  $\tau$ is the ion-memory decoherence time, as now described. Those modelled states are obtained by the two-qubit map

\begin{equation}
\tilde{\rho}_{Aa}(t, \tau) =  p(t)\times \tilde{\rho}_0 + (1-p(t)) \left( S_z^{\dagger}\tilde{\rho}_0 S_z\right).
\label{gaussian}
\end{equation}

Here $S_z$ is defined after Eqn. \ref{eq:depol}, 
$p(t) = (1-e^{-\frac{t^2}{\tau^2}})/2$, $\tau$ is the characteristic Gaussian decoherence time, $\tilde{\rho}_0$ is the initial ion-photon state stored in memory, the Hilbert space order is $H_{ion}\otimes H_{photon}$. For the initial state $\tilde{\rho}_0$, the tomographically reconstructed on-photon state at step $k=0$ is used, that is $\tilde{\rho}_0=\rho_{Aa}(k=0)$. 
The map of Equation \ref{gaussian} leaves the photon qubit unchanged and is therefore referred to as a one-qubit map in the main paper.

The fidelities of the modelled states $\tilde{\rho}_{Aa}(t, \tau)$ are calculated with respect to the maximally-entangled state that is nearest to $\tilde{\rho_0}$.
A fit of the modelled fidelities to the ones from data, optimised over $\tau$, yields the red solid line in Figure 4 of the main text. The model is seen to describe the measured fidelities well, to within statistical precision, and yields $\tau =62\pm{3}$ ms.

\subsubsection{Model of the two-photon states delivered by the repeater protocol}    
\label{sec:DBSM_sim}

The previous subsection established  a model for the ion-photon states stored in memory.  
To recap, that model captures the decoherence of the ion-memory observed in the ion-photon states measured without \SI{25}{km} fiber spools and telecom conversion. 
In this subsection, we describe the extension of that model to predict the final two-photon telecom states delivered by the repeater protocol, assuming no other imperfections.  The purpose of developing this extension is to determine to what extent ion-memory decoherence captures all the imperfections observed in the repeater protocol. 

The model of the two-photon states delivered by the repeater protocol is developed as follows. 
First, we obtain a four qubit state which is a tensor product of the two ion-photon states established at step $k$ of Loop 2.
Specifically, for the ion-photon state stored in memory, the modelled states $\tilde{\rho}_{Aa}(t, \tau = 62ms)$ are used in all cases. 
For the ion-photon states not stored in memory, the measured states $\rho_{Aa}(k=0)$ are used in all cases. 
Second, the unitary rotation $U=\exp(-i\sigma_x^1\sigma_x^2 \pi/4)$, describing the MS operation as introduced in \refsec{DBSM}, is applied numerically to the ion qubits. 
Third, the obtained 4-qubit state is projected onto a logical ion state measurement outcome.
Fourth, the photon-photon state expected for that ion-outcome is obtained by tracing out the ions.
Fifth, a perfect feed-forward is simulated resulting in a single predicted Bell-like state.
All other parts of the repeater protocol, and simulated feedforward, are modelled without imperfection.

The predicted two-photon Bell state fidelities, established between the remote nodes at step $k$ of Loop 2, are plotted as a dashed black line in Figure 4. Finally, the modelled two-photon states at each attempt ($k$) are added up as a mixture, weighted by the probability with which they occurred in the repeater experiment (Figure 3b), yielding a predicted photonic Bell-state state fidelity for the repeater protocol of 0.813(7). The uncertainty in this fidelity value comes from the uncertainty in $\tau$ and the statistical uncertainty in the tomographically-reconstructed ion-photon state $\rho_{Aa}(k=0)$. The latter is quantified via the Monte Carlo approach in an analogous way to the method in \refsec{sec:probs_and_MC} for ion-ion states, yielding a set of  500 noisy ion-photon states.   
We calculate 500 fidelity values by applying the model to the noisy states for the case $\tau = 62+3$ ms.  
 We calculate another 500 fidelity  values for the case $\tau = 62-3$ ms. 
 We define the error in the fidelity as the standard deviation of the 1000 fidelity values obtained. 
By comparing that prediction with the experimentally obtained value   $F_{ff}=0.72^{+2}_{-2}$ (see \refsec{sec:feedforward}), we conclude that our model captures the dominant sources of infidelity in the repeater protocol: decoherence of the ion-memories.

\subsubsection{Investigation into origins of ion-memory decoherence}

We study the sources of the decoherence of the memory qubit by performing Ramsey-type experiments. 
Beginning with the state $\ket{S,S}$, these experiments consist of two effective $\pi/2$ pulses on the $S/S^\prime$ qubit transition with a waiting time $t_{wait}$ in between. 
Each effective $\pi/2$ pulse consists of two laser pulses at \SI{729}{nm} that act equally on both ions: a $\pi/2$-pulse on the $\ket{S}$ to $\ket{D'}$ transition and a $\pi-$pulse on $\ket{S'}$ to $\ket{D'}$ transition. By scanning the optical phase $\phi$ of the second $\pi/2$ pulse, and measuring ion's excitation probability $P_e$, we obtain the standard Ramsey fringe pattern. The amplitude of the fringe $C$ is obtained as a fit parameter given the fit function $P_e = C\sin^2(\phi)$. 
We model decoherence of the ion-qubit, during the wait time $t_{wait}$, using the one-qubit map obtained after eliminating the photon from the two-qubit map of Equation \ref{gaussian}. The obtained one-qubit map is simply single-qubit dephasing with a Gaussian temporal profile. It is straightforward to show that, under the action of this map, the Ramsey amplitude decays as $C=C_0 e^{-\frac{(t_{wait})^2}{\tau^2}}$ where $C_0$ is set to 0.99 according to an independent measurement of the Ramsey amplitude.

In the first Ramsey experiment we set $t_{wait}=$ 66ms and, during that time, execute the Loop 2 sequence of the repeater protocol in the case where ion A encodes a qubit. That is, spin echoes are performed on ion A (during $t_{wait}$) while 195 photon generation attempts are performed on neighbouring ion B. 
The obtained Ramsey amplitude on ion A is $C=0.27(4)$ corresponding to a decoherence time of $\tau=$59(1) ms. This coherence time is consistent with the 62(3) ms obtained from the decay of the ion-photon states in Figure 4 of the main text. 
The second Ramsey experiment is identical to the first except no photon generation pulses are applied to ion B. The obtained Ramsey amplitude on ion A is $C=0.67(3)$ corresponding to a decoherence time of $\tau=$108(1) ms. One sees that the coherence time of the ion-A memory approximately halves in the  presence of the repeated photon generation attempts (Raman laser pulses) on ion B. Next we turn to identifying the physical origin of the additional memory decoherence (additional beyond the ambient decoherence without nearby laser pulses) in the presence of Raman pulses on the neighbouring ion.

The third Ramsey experiment is identical to the first---full Loop 2 sequence---except the \SI{854}{nm} repump laser is removed thus leaving ion B in the $D_{j=5/2}$ manifold after the first photon generation pulse. In that manifold, ion B does not emit or absorb any photons due to photon generation pulses.  The obtained Ramsey amplitude is indistinguishable from the highest one obtained without photon generation pulses (the second Ramsey experiment, where $C=0.67(3)$). From this one can exclude `addressing' errors in the focused Raman beam as a cause of additional decoherence: the focused 195 Raman laser pulses aimed at ion B do not directly cause observable decoherence on the ion-A memory e.g., due to resonant or off-resonant effects. Moreover, it is clear that photon absorption and emission from ion B is necessary to cause the additional ion-A memory decoherence.

The fourth Ramsey experiment is identical to the first---full Loop 2 sequence---except the photon generation process on ion B is set off-resonant, such that no \SI{854}{nm} cavity photons are generated. In that case, the Raman laser pulse in each attempt can still cause off-resonant excitation of the electron in ion B to the $P_{j=3/2}$ manifold causing photon scattering primarily at \SI{393}{nm}. The obtained Ramsey amplitude is indistinguishable from the reduced one obtained from the first Ramsey experiment ($C=0.27(4)$). From this, one can conclude that the scattering of \SI{393}{nm} photons by ion B, during Raman laser pulses aimed at ion B, is the dominant cause of additional decoherence on the ion-A memory, beyond the background level.  \\

We expect that the most likely mechanism through which \SI{393}{nm} photon scattering compromises the neighbouring ion-memory is as follows: photon scattering causes heating of the joint motional modes which in turn cause imperfections in the spin-echoes used on the memory ion. In the following, we present results of a simplified model of a Ramsey experiment with imperfect spin echos to substantiate this expectation. The model, presented in Section \ref{sec:spin_echo_model}, is simplified in several ways. First, the model considers only imperfections in the spin echoes, not any other decoherence mechanism like e.g., fluctuations in ambient magnetic fields, for which the spin echoes are there to overcome. Second, the model considers motional mode phonon number occupancies that are elevated but fixed throughout the spin-echo sequence (\ref{sec:temp}), whereas in reality those numbers are dynamically evolving\footnote{Modelling the spin-echo sequence with the dynamical temperature is beyond this work. The temperature is predicted to grow linearly with the number of the photon scattering events.} during Loop 2. Nevertheless, the model, which is based on independently-calibrated average phonon numbers, predicts a significant reduction in the Ramsey visibility due to the temperatures expected in our repeater experiment.

Here we provide results of the Ramsey spin-echo model for two cases. The first case aims to model the second Ramsey experiment, in which the full Loop 2 sequence is implemented without photon generation pulses. Specifically, the model employs the coldest motional states expected in the beginning of Loop 2.  The second case aims to model the first Ramsey experiment, in which the full Loop 2 sequence is implemented with photon generation pulses. Specifically, the model employs the motional states expected in the middle of Loop 2 (the average temperature in Loop 2). 
The model returns a Ramsey visibility of 0.92 in the first case, and 0.67 in the second case: a significant drop due to spin-echo imperfections caused by the elevated temperature in the latter case. 
To recap, the corresponding Ramsey experiments yielded 0.67(3) and 0.27(4). 
In conclusion, the model shows that a significant contribution to the ion-memory decoherence caused by photon generation attempts on the neighbouring ion is caused by the subsequent heating of the ion string due to photon scattering which in turn compromises the spin-echo pulses.

Note that the rise in the ion string temperature compromises the \SI{729}{nm} $\pi$-pulses used in the composite spin-echo sequence in two ways. First, it decreases the maximum transition probability, second it reduces the effective Rabi frequency. The second could be corrected by re-calibration. For both the model, and in the repeater experiment, we optimised the calibration of the \SI{729}{nm} $\pi$-pulses for the thermal states at the beginning of Loop 2. If we allow for an additional 1\% error in the calibration of each of the $\pi$-pulses duration, then the calculated amplitude of the Ramsey fringe in case 2 of 0.73 drops down to 0.62.

\subsubsection{Model of imperfect spin-echos}
\label{sec:spin_echo_model}

As described in section \ref{sec:spinecho}, each spin echo is composed of a sequence of three \SI{729}{nm} $\pi$ pulses that operate on the three relevant atomic levels  $\ket{S}, \ket{S'}, \ket{D'}$, followed by an incoherent repumping of the $\ket{D'}$ level's population (as realised in the experiment with an \SI{854}{nm} repump pulse). 
For our model of imperfect spin echos, we write the state of the three levels system as three level (qutrit) vector, e.g., $[1,0,0]$ for the $\ket{S}$ state.
The spin-echo sequence is modelled for a finite string temperature in the following way. Each composite spin echo is described as a product of three qutrit operators, one for each of the \SI{729}{nm} pulses.  For example, for the pulse that couples the states $\ket{D'}/\ket{S'}$, the qutrit operator is: 
\begin{center}
$\left( \begin{array}{ccc}
1 & 0 & 0 \\ 
0 & \cos(\Omega T/2) & ie^{i\phi}\sin(\Omega T/2) \\
0 & ie^{-i\phi}\sin(\Omega T/2) &  \cos(\Omega T/2) \\ 
\end{array}\right)$,
\end{center}
where $\Omega$ is the laser drive Rabi frequency and $T$ is the pulse duration. The other operators are constructed by analogy. The modelled imperfection is the reduction of the drive laser Rabi frequency, by the same amount for all three operators, due to coupling to the motional sidebands of the two-ion string. Specifically,  $\Omega = \Omega_0 (1-\eta_1^2n_1) (1-\eta_2^2n_2) (1-\eta_3^2n_3) (1-\eta_4^2n_4)$ where $\eta_i$ and $n_i$ are the Lamb-Dicke parameter and a phonon occupation number of the $i^{th}$ motional mode  (in order: radial COM, radial rocking, axial COM and axial stretch modes), respectively.

The three qutrit operators act sequentially, starting from the state created by the first $\pi/2$-pulse of a Ramsey sequence: $\ket{+} = 1/\sqrt(2)[1,1,0]$, which we describe by the density matrix $\rho =\ket{+}\bra{+} $. 
Each composite spin echo is followed by incoherent repumping described by the map: $\rho \rightarrow \rho_{22}\times S_{mixed} +(1-\rho_{22})\rho^\prime/Tr(\rho^\prime)$ where $\rho_{22}$ is the element of the density matrix representing the population of the $\ket{D'}$ state, $S_{mixed}$ is the qutrit density matrix representing an equal mixture of $\ket{S}$ and $\ket{S'}$, and $\rho^\prime$ is the same as $\rho$ except that $\rho^\prime_{22} = 0$. 

Using this sequence of operations for each composite spin echo, and for a fixed choice phonon numbers (n1, n2, n3, n4), we compute the density matrix after 40 sequential composite spin echos. That number of composite spin echoes is the same as implemented during Loop 2 of the repeater protocol when 200 photon generation attempts occur (a spin-echo is done every 5th attempt). That process is repeated for all possible combinations of choice of phonon occupation numbers, where each is iterated from 0 up to 40, resulting in $40^{4}$ different combinations and the same number of density matrices.   
Finally, we add up those density matrices in a statistically mixture, where each is weighted by the product of the individual mode's thermal state distributions. The temperatures (mean phnon numbers) of those modes is set either to be the values estimated for the beginning of Loop 2, or for the end of Loop 2. 
Finally, in either case, we calculate the expected amplitude of the Ramsey fringe $C$ of the mixed state.

\subsubsection{Heating of the ion string due to photon generation attempts}
\label{sec:temp}

The photon generation process in the repeater protocol causes heating of the ion string due to photon recoil. For example, a \SI{393}{nm} photon is absorbed from the Raman laser whenever a cavity-photon is generated, causing a recoil kick to the ion in the direction of the Raman laser. Moreover, recoil kicks accompany unwanted spontaneous scattering events from the $\ket{P_{3/2}}$ state. During each of those scattering events, a 393 nm photon is absorbed from the Raman laser and either a 393 nm or a 854 nm photon is emitted into free-space.

The two-ion string used in the repeater protocol has six modes of motions: Axial ($ax$) center of mass ($COM$) and stretch ($STR$) modes and radial ($rad$) center of mass and rocking ($ROCK$) modes in two orthogonal directions (1,2). 
Our best estimate of the corresponding six motional frequencies in the repeater experiment are $\{f_{ax}^{COM}, f_{ax}^{STR}, f_{rad, 1}^{COM}, f_{rad, 1}^{ROCK},f_{rad, 2}^{COM}, f_{rad, 2}^{ROCK}\}$ 
$=\{0.963(5), 1.668(9), 2.18(1), 1.956(9), 2.13(1), 1.900(9)\}$ MHz. Here, all COM frequencies were measured via 729 nm spectroscopy while the others are predicted numerically by solving the associated equations of motion. 
In the following, he radial COM and Rocking modes are considered to be degenerate: the average frequencies of those given in the vector are used.

Before presenting our experiment and model for ion-heating, we first present the results. The results provide estimates for the following set of four mean phonon occupation numbers at specific points during the repeater protocol: $\{n_{ax}^{COM}, n_{ax}^{STR}, n_{rad}^{COM}, n_{rad}^{ROCK}\}$. 
For the conditions of the beginning of Loop 2 of the protocol we estimate mean phonon occupation numbers $\{0, 11, 8, 9\}$.
For the conditions at the end of Loop 2 of the protocol---after {210} photon generation attempts--- we estimate mean phonon occupation numbers $\{9.2, 16, 29, 34\}$.  
Those values are used for further calculations in \refsec{sec:spin_echo_model} and \refsec{sec:WP_simulation}. 
How those results are obtained is now described. 

At the beginning of Loop 2 we expect that the all motional modes of the ion string, except the axial COM, are at temperatures achieved by Doppler cooling. The axial COM  should be close to the ground state, having undergone sideband-resolved ground state cooling within the last five photon generation attempts in Loop 1.  
To estimate the corresponding average phonon occupation numbers we conduct an experiment in which a single ion is trapped with the same COM frequencies as two-ion string in the repeater protocol.  Doppler cooling followed by sideband-resolved ground state cooling is performed. The mean phonon number in the axial COM is measured to be $n_{ax}^{COM} =  0.12(7)$, via the method of fitting measured \SI{729}{nm} Rabi flops on the sidebands to the predictions of a thermal motional state (see e.g. \cite{Kirchmairthesis}). Since that value is much smaller than the phonon numbers estimated in the other modes, we use $n_{ax}^{COM} =  0$ for simplicity. 
The mean phonon number in the radial COM is measured to be $n_{rad}^{COM} = 8(0.5)$, determined by the method of fitting measured \SI{729}{nm} Rabi flops on the carrier to the predictions of a thermal motional state and using a \SI{729}{nm} beam that is perpendicular to the axial direction. 
Finally, all the mean occupation numbers of the two-ion string, except $n_{ax}^{COM}$, are then calculated under the assumption they each contain the same mean thermal energy after Doppler cooling as the value of the radial COM mode measured for the single ion. Specifically, the values given in the beginning of this section and used elsewhere in the manuscript are derived from $n_{rad}^{COM} = 8$.

We determine the phonon occupation numbers at the end of Loop 2 in two steps. First, we experimentally measure the heating rate of the axial COM due to photon generation attempts. Second, we construct a model of that heating that allows us to estimate the heating rate of other motional modes due to photon generation attempts. 

The experiment for measuring the heating of the axial COM mode in short was the following. 
Two ions are trapped with the same confinement as in the repeater protocol. The ions are Doppler cooled then sideband cooled, initialising them into motional states similar to the ones at the beginning of Loop 2 in the repeater protocol. 
Next, the following sequence of pulses is repeated a variable number of times: 
a bichromatic Raman laser pulse is applied to ion B, followed by a global optical pumping pulse at \SI{393}{nm}\footnote{In the repeater experiment there was no optical pumping pulse in Loop 2. Unfortunately, directly comparable data of ion temperature was not taken for the case of no optical pumping, however it was found that the optical pumping pulse had negligible effect on heating for a similar experiment with a single ion.} 
in parallel with global repumping pulses at \SI{854}{nm} and \SI{866}{nm}. 
That pulse sequence corresponds to a photon generation attempt on ion B and is equivalent to an attempt in the Loop 2 sequence of the repeater experiment. For reference, the power in the Raman beam (before being split into the bichromatic components) was $\Omega_{393} = (2\pi)45.5(5)$ MHz, which is comparable to the value in the repeater experiment of $\Omega_{393} = (2\pi) 41.0(5)$ MHz. In both experiments, each Raman pulse was applied for $50\mu s$.

The temperature of the axial COM mode of the ion string was measured as a function of the number of attempts on ion B.
The method was to drive the corresponding red and blue sideband Rabi flops after global optical pumping and extract the mean phonon number from fits of the predicted dynamics for a thermal state, and also from their ratio. The experiment was performed for up to 60 attempts on ion B and showed a linear increase in the mean phonon occupation number with the number of attempts. The result is that the axial COM mode of the two-ion string heats by 0.043(4) phonons per photon generation attempt on ion B. 

We numerically model the heating of the ion string caused by the photon generation pulses due to absorption and spontaneous emission events. Specifically, we use displacement operators acting onto the four harmonic oscillators with initial thermal states as for the beginning of Loop 2. The number of spontaneously scattered \SI{393}{nm} photons per photon generation attempt is a parameter of the model that we do not directly measure. Instead, we use the measured heating rate for the axial COM mode of 0.043(4) phonons per photon generation attempt to calibrate the model. 
Finally, the calibrated model is used to obtain the mean photon occupation numbers in all four motional modes after the 210 photon generation attempts of Loop 2,  that are provided at the beginning of this section.
Those numbers are calculated from a model calibrated with exactly 0.043 phonons per pulse heating and no error propagation.

\subsection{Numerical simulation of the cavity photon generation efficiency}
\label{sec:WP_simulation}
Numerical simulations were performed to obtain an estimation for the photon generation efficiency. Specifically, the master-equation model of the laser-atom-cavity system is used from \cite{SchuppPRX2021}. 
The model parameters include the experimental geometry as described in \refsec{sec:setup}. 

The strength of the bichromatic drive $\Omega = \sqrt{\Omega_{1}^2 +\Omega_{2}^2}$ is set such that the model predicts the same AC Stark shift of the Raman transition as the one measured in the repeater experiment (0.73 MHz, see \refsec{Photon_generation_method}). 
The ratio between the powers in the two components of the bichromatic drive is set at the value for which the model predicts equal probabilities for the generation of the H and V polarized photons.
The resultant drive field strengths correspond to $\Omega_{1} = (2\pi) 25.997$ MHz and $\Omega_{2} = (2\pi) 31.704$ MHz when expressed in the terms of the Rabi frequencies for the $|4^2\text{S}_{1/2}, m_j = -1/2\rangle$ to $|4^2\text{P}_{3/2}, m_j = -3/2\rangle$ transition driven with $\sigma^-$-polarization and laser beam direction parallel to the magnetic field. The laser beam in the experiment has a different direction and polarization, as described in Section \refsec{sec:setup} and the correction is taken into account with appropriate geometrical factors and Clebsch-Gordan coefficients.  

The effective coupling strength of the photon generation process (BCMRT) is $\Omega_{eff}\propto g\Omega$, where g is the ion-cavity coupling strength \cite{SchuppPRX2021}.  To account for various mechanisms that reduce the effective coupling strength we use the expression $g\Omega= \alpha_1\alpha_2x \beta g_{max} \Omega$, where $g_{max} = (2\pi)1.53$ MHz is the maximum ion-cavity coupling strength in our system \cite{SchuppPRX2021} and $\alpha_1,\alpha_2, x$ and $\beta$ are all numbers between 0 and 1, that each represent a reduction factor. 

The factors $\alpha_1$ and $\alpha_2$ quantify the expected reduction due to the coupling of the Raman laser to the radial COM and Stretch modes, respectively\footnote{The dominant effect of being outside the ground state on cavity-photon generation in our system is due to changes in the coupling of the drive laser to the ion.}. 
Values of $\alpha_1 =0.97$ and $\alpha_2 = 0.96$ are used, calculated using the method described in Appendix C of \cite{SchuppPRX2021}. Here, we use the mean phonon occupations of the radial modes estimated at the beginning of Loop 2: the lowest temperatures considered \ref{sec:temp}. 
The factor $x$ quantifies the reduction in ion-cavity coupling strength due to the ions not being positioned exactly in the cavity vacuum standing wave, as described in \refsec{sec:trapandconversion}. A value of $x=0.935$ is used for both ions. 
The factor $\beta$ quantifies any other reduction in effective coupling strength of the photon generation process.

For $\beta=1$, the model yields a probability for generating a photon in the cavity mode of 0.61, after 40 $\mu$s of the photon generation Raman laser pulse.  
However, the modelled and measured single photon wavepackets from Loop 1 (measured wavepackets are shown in Figure 3a) exhibit a statistically significant mismatch. 
Additional modelling shows that the measured photon wavepackets can be well reproduced by the model by setting $\beta=0.79$. In that case, the model predicts a probability for photon generation in the cavity of 0.52. 
A possible cause of such a reduction is an increase in cavity length jitter compared to that considered in \cite{SchuppPRX2021}. In the current experiment we significantly reduced the power of the \SI{806}{nm} laser light sent into the cavity for length stabilisation in order to reduce the AC Stark shifts that it causes, which could explain a poorer cavity-length stabilisation.

\section{Future systems}
\label{sec:future}

\subsection{Model of repeater chains: stored entanglement over 800 km}
\label{sec:1000km}

In the main paper, it is written that {by modelling a chain of 17 enhanced ion nodes across an 800 km fiber, the predicted average time to establish heralded entanglement between ions in the repeater nodes at either end ($T_{tot}$) is 0.07s. Moreover, the predicted Bell state fidelity obtained is 0.61}. The parameters of those enhanced ion nodes are given in the main text. Details on the underlying calculations are presented in this section.\\

{{We consider the envisioned repeater chain setup modelled in \cite{Sangouard2009},
in which entanglement is established between neighbouring ion-nodes (each containing two ion-qubits) by interfering and detecting two photons (one produced by each ion-node) at intermediate photonic stations between the nodes. Those intermediate photonic stations could be the photonic nodes in our present experiment (Figure 2), where the photon from a remote ion-node enters through the empty input port of the polarising beam splitter.} Two fibers with an intermediate photonic station and an ion-node at each end is henceforth called a `link'.} 
The detection of two photons at the intermediate station realizes a Bell state measurement on the photons resulting in heralded entanglement of the two ions at the ends of the link. 
{Once entanglement has been established across links,  one ion-qubit Bell state measurement (each referred to as DBSM in the main text) is carried out in each ion-node to swap the entanglement out to between the end-nodes.} 
\\

\noindent \textbf{800 km Rates.}

{\noindent As explained in the following, we use Equation 5 of \cite{Sangouard2009} to calculate the average time for heralded entanglement distribution between the ion end-nodes ($T_{tot}$)} 
{in a repeater chain that spans over 800 km of optical fiber. Specifically, the chain consists of a total of 17 two-ion repeater nodes, of which 15 are used as repeaters and the two outer ones as the end nodes, forming 16 concatenated links. Each link contains a total of  $L_0 = 50$ km of fiber with an intermediate photonic station in the middle.} 
{The key parameter for each repeater node in this model is $P^0_{link}$: the combined single telecom photon generation and detection probability (at the photonic station) for each node excluding fiber losses. The fidelities of various operations are not involved in the model for $T_{tot}$. Note that Equation. 5 of \cite{Sangouard2009} is an approximation, which points to the fact that it is correct up to a factor of $1.5$, as explained in that work. In this present work, we ignore this small approximation by replacing the approximation sign with and equals sign.}

{Equation. 5 of \cite{Sangouard2009} assumes that the entanglement distribution attempt rate is limited by the sum of the travel times of the photon to the photonic station and classical return signal heralding the photon detection. That total time is $L_0/c'$, where $L_0$ is the distance between the neighbouring repeater nodes and $c'$ is the propagation speed of both the photon and classical signal. However, in our present experiment, the time taken to generate each photon ($T_0$) is a significant fraction of the photon and signal travel time and therefore should be included. We include $T_0$ by making the following simple substitution into Equation. 5 of \cite{Sangouard2009}:  $L_0/c'\rightarrow L_0/c'+T_0$.  The modified and final equation using the notation established in this paper is}

\begin{equation}
T_{tot} = \left(\frac{L_0}{c'}+T_0\right)\frac{3^n}{2^{n-1} (P^0_{link})^2 \eta^2_t}. 
\end{equation}

{\noindent Here, $\eta_t$ is the half-link transmission probability introduced before Eq. 5 of \cite{Sangouard2009} which, in the notation of our manuscript, is expressed as $\eta_t  = \sqrt{\eta} = 10^{-\gamma L_0/2}$.} 
{The symbol $n$ is the number of `entangling nesting levels' required to swap entanglement from the links all the way out to the end nodes \cite{Sangouard2009}. That is, $n$ is the number of steps required to swap entanglement from being first between neighbouring nodes (across one link), then to being between nodes separated by two links, then to being between nodes separated by four links and so on to all until the ends are reached. For our scenario with 16 links, $n=4$ levels are required. } 
We use $T_0 = 175 \mu$s, corresponding to the photon generation time in Loop 1 for the experiment in Figure 3, where the time to generate two photons is taken, given the sequential generation process.

{Finally, and in summary, using $P_0=0.21$, $T_0 = $175 $\mu$s, $L_0=$\SI{50}{km}, $c'=2\times10^8$ m/s,  $n=4$ and $\gamma = 0.0173$ km$^{-1}$ we solve equation S13 and obtain an expected average entanglement distribution time over the 800 km chain of $T_{tot}=0.71$ s. The inverse of $T_{tot}$ is 1.4 Hz: the average rate of entanglement distribution.}

In order to achieve similar success rates for a repeaterless telecom 800 km fiber link, the attempt rate would have to exceed the $10^{14}$ Hz which is the carrier optical frequency of the telecom photons. Achieving stored entanglement in the latter case would require deep  multimoding ($\sim 10^{11}$ modes).\\

\noindent \textbf{800 km Fidelity.}

{\noindent We develop a simple numerical model to predict the Bell state fidelity of the heralded state established between the ends of the \SI{800}{km} repeater chain. 
The model assumes that the memory decoherence time in each node is much longer than $T_{tot}$, such that it can be ignored in the entanglement distribution process. }
{The following imperfections are considered in the model: 1. the imperfect ion-photon state fidelities ($F_0$) 2. the imperfect fidelity of the DBSM ($F_{swap}^{ions}$) and 3. the imperfect fidelity of the Bell state measurement on the photons ($F^{photons}_{swap}$). We use the values for the aforementioned parameters that are given in the main text for the enhanced node, for each node in the chain.  The four steps of model is now described.}

First, the initial ion-photon state {density matrix} $\rho_{i-ph}$ produced by each node, with Bell state fidelity $F_0$, is obtained via the dephasing model $\rho_{i-ph} = F_0\ket{\Psi^{-}}\bra{\Psi^{-}}+(1-F_0)(\ket{\Psi^{+}}\bra{\Psi^{+}})$, where $\ket{\Psi^{\pm}}$ are the corresponding Bell states. 

Second, we model the ion-ion state $\rho_{ii}$ between each pair of neighbouring nodes as  
a result of an imperfect photonic Bell state measurement on the two relevant ion-photon states (each $\rho_{i-ph}$).
{The modelled imperfection here stems from the partial distinguishability of the photons produced at different nodes: The fidelity of the photonic Bell state measurement $F^{photons}_{swap}$ for ion-cavity systems was shown to be predominantly defined by the two photon interference visibility $V$ as $F^{photons}_{swap} = (1+V)/2$ \cite{Krutyanskiy:2020pra, ionion}, where $0\leq V\leq 1$. To model this effect we follow the approach described in the supplementary material of \cite{Quraishi19_swapping}}.
Specifically, the ion-ion states for the case of perfect ion-photon states $\ket{\Psi^{-}}$ or $\ket{\Psi^{+}}$ are provided by equation (S29) of \cite{Quraishi19_swapping}. 
To obtain (S29), the authors applied the measurement operator of Equation (S28) to those perfect states.
We obtain the swapped ion-ion states $\rho_{ii}$ for our case of imperfect ion-photon states by applying operator (S28) of \cite{Quraishi19_swapping} to the mixed input states $\rho_{i-ph}$.
For example, in the case where the $\ket{\psi^-}\bra{\psi^-}$ outcome of the photonic Bell state measurement is obtained, the result is

\begin{equation}
 \begin{multlined}
\rho_{ii} = F\ket{\Psi^{-}}\bra{\Psi^{-}}+(1-F)\ket{\Psi^{+}}\bra{\Psi^{+}},\\
\label{eq:rho_ii}
 \end{multlined}
\end{equation}   

where

\begin{equation}
F = \frac{(1+V(1-2F_0)^2)}{2}
\end{equation}

and $V$ is the two-photon interference visibility and $F_0$ is the ion-photon Bell state fidelity produced by each node. 
In the case of $F_0=1$, a state $\rho_{ii}$ is obtained with Bell fidelity $F=F^{photons}_{swap} = (1+V)/2$ in agreement with Equation (S32) of \cite{Quraishi19_swapping}  and with \cite{Krutyanskiy:2020pra}.

{Third, starting with the ion-ion states between neighbouring nodes $\rho_{ii}$, we perform, numerically, four ($n=4$) `nesting levels' of the DBSM at the ion nodes, where the output states $\rho^m_{ii}$ of the level $m$  are used as the input states for the next level}. Those four layers correspond to realising the 16 links in the chain. 
At each layer, the DBSM are done first by applying a perfect operation, followed by a depolarising channel on the two-qubit output states $\rho^m_{ii}$. 
Specifically, $M^{depol}(\rho^m_{ii}, F_{swap}^{ions})$ using the map of Eq. \ref{eq:depol}. 
{Fourth and finally}, we calculate the fidelity between the output two-qubit state of the final {leve}l ---encoded in one ion at each of the nodes at the ends of the chain --- and the nearest maximally entangled {pure} state. The result is $0.61$, which is above the separable state threshold of $0.5$. 

\subsection{Model of secret key rate between photonic endnodes.}
\label{sec:SKR}

{An application of quantum networks is to establish a secret key between two users \cite{Scarani09}. 
Theoretical bounds have been derived  for the rate at which a secret key could be established between two users without using quantum repeaters in between: that is, by direct transmission through an optical channel \cite{Pirandola2017}. 
With a single quantum repeater in the middle of the channel, one could overcome this bound (see e.g., \cite{Weinfurter2020_theory_repeater}).  
In this section, we calculate the performance enhancements that would be required in order for \emph{one} of our ion repeater nodes to establish a secret key between photonic endnodes (users) at a rate greater than the theoretical upper bound for any repeaterless link with attempt rates limited by the photon travel time. Overcoming that bound over tens of kilometres is an open benchmark \cite{Weinfurter2020_theory_repeater, Wehner19, Langenfeld_Rempe_21,Bhaskar_Lukin2020}. 
This calculation is not presented in the main text, but serves as an intermediate future target for performance, before repeater chains. }

\begin{figure}[h!]
	\vspace{0mm}
	\begin{center}
        \includegraphics[width=0.99\columnwidth]{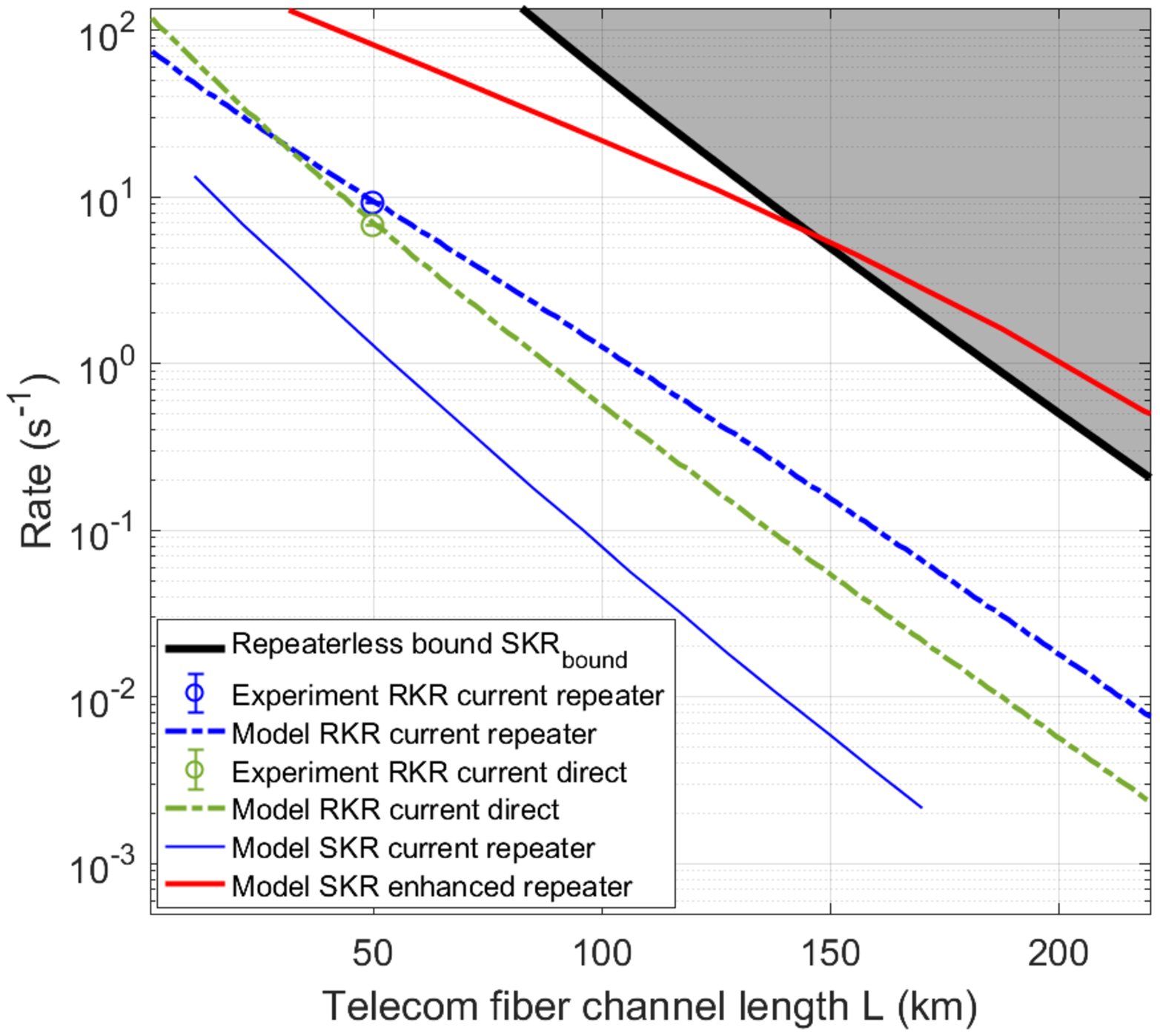}
		\caption{
			\textbf{Secret key rate (SKR) and success rate (raw key rate, RKR) for the current and enhanced nodes.} The situation considered is a single repeater node distributing a photonic key. The black solid line shows the SKR bound $\mathrm{SKR}_{bound} = -\log_2(1-\eta)\times c/L$ ($\eta$ is the fiber transmission, see \refsec{sec:SKR}). The
SKR higher than the black line (marked as shaded area) is not reachable without repeaters if the attempts rate  is limited by the photon-travel-time ($L/c$). For exact conditions of applicability of that bound  see \cite{Pirandola2017}, where a term `channel uses' is used instead of `attempts'.	Symbols show the success rates (RKR) achieved in our experiment for comparing the repeater and direct configurations (see the legend). Dashed lines show modeled RKRs for independently measured parameters of these experiments (see \refsec{sec:numeric_rates}). Solid lines show modeled SKRs for the parameters of current and enhanced repeater nodes  (see \refsec{sec:SKR}).  
						}
		\label{fig:rates}
		\vspace{2mm}
	\end{center}
\end{figure}

In summary, the secret key rate {SKR} achievable given modest system enhancements is calculated by adopting the model of \cite{Weinfurter2020_theory_repeater} for a repeater node between two photonic endnodes.
The relevant parameters of the model are: memory coherence time ($\tau$);  photon detection probability at each endnode after removing losses in the channel ($P^{0}_{link}$,  \refsec{0km}); fidelity of each ion-photon entangled pair ($F_{0}$); fidelity of the entanglement swapping operation ($F_{swap}^{ions}$); the total length of the fiber between the photonic nodes ($L$); the spatial attenuation rate of the fiber ($\gamma$); the time for photon generation and reinitialisation ($T_0$); 
the time for the DBSM and feedforward ($T_{swap}$). 
Here we use, as before, that the probability of a photon transmission through a fiber $\eta$ is defined as $\eta = 10^{-\gamma L}$.
We write the model parameters as a vector $[P^0_{link},\tau, F_{0}, F_{swap}^{ions}, L,\gamma, T_0, T_{swap}]$ and apply it to the six state protocol \cite{Scarani09, Wehner19}. 
We consider an asymetric six-state protocol for which the SKR is defined as  $\mathrm{SKR = RKR\times SKF}$ \cite{Scarani09, Wehner19}  Where $\mathrm{RKR}$ is the 
raw key rate (RKR) and the $\mathrm{SKF}$ is  the secret key fraction (SKF) and a distributed entangled (photon) pair can be used to get a raw secret key bit \cite{Scarani09}.
Various quantum key distribution protocols tend to the same SKR scaling with $L$ in the limit of small channel transmission probability ($\eta \rightarrow 0$)  \cite{Weinfurter2020_theory_repeater,Pirandola2017} thus choosing a particular protocol is not crucial for the conclusions, provided that  $\eta$ is small enough. 
The calculation is done following \cite{Weinfurter2020_theory_repeater} and contains the following three stages. (Note, in the following we refer to the attempts to distributed a photon in Loop 2 of the repeater protocol as `memory attempts', since those attempt occurs while established remote entanglement is stored in memory).  

In stage one, the photon-photon density matrix $\rho_k$ of the state distributed in the $k-$th memory attempt of Loop 2 is calculated. The $k-$th memory attempt is associated with the waiting time $t = k\times T$ where  $T = T_0+L/c$  and $c$ is the photon propagation speed in the fiber as before. We calculate $\rho_k$ as $\rho(\tau, t = k\times T)$ of the deoherence model of Equation \ref{gaussian}   (\refsec{sec:memory}) with the Gaussian decoherence parameter $\tau$. 
That model requires the initial ion-photon state (at $t = 0$), which we here denote as $\rho_{t=0}$. 
We construct $\rho_{t=0}$ by applying a depolarizing channel $M^{depol}$ to the ideal ion-photon Bell state $|\Phi^+\rangle \langle \Phi^+|$  yielding an output state with Bell-fidelity $F_0$, specifically $\rho_{t=0} = M^{depol}(|\Phi^+\rangle \langle \Phi^+|, F_0)$, where \\
\begin{equation}
M^{depol}(\rho, F) = F\rho+\frac{1-F}{3}(S_z\rho S_z+S_y\rho S_y+S_x\rho S_x),  
\label{eq:depol}
\end{equation}
and  $S_i \in [I\otimes I, I \otimes \sigma_z, I \otimes \sigma_x, I \otimes \sigma_y]$ as introduced in \refsec{sec:feedforward}.
We then apply same depolarizing channel to the state $\rho_k$ to take into account for the finite fidelity $F_{swap}^{ions}$ of the DBSM, i.e we calculate $\rho^\prime_k = M^{depol}(\rho_k, F_{swap}^{ions})$. 
Then we derive the SKF for that state $\mathrm {SKF}(\rho^\prime_k)$ by calculating quantum bit error rates (see e.g. \cite{Wehner19}) and  applying eq. G21 of \cite{Wehner19}. At the end of stage one we find the cut-off number of steps  $K$ such that for $k \leq K$ it holds that $\mathrm{SKF}(\rho_K) \geq 0.1$. The repeater node could provide an even higher SKR then our estimation if the latter fixed condition is replaced by an optimization procedure as it is done in  \cite{Weinfurter2020_theory_repeater}.

In stage two, we find the probability $P_k$ of successful state distribution at the memory attempt $k$ and the average number of attempts $\overline n$ for successeful distribution. Both are found using  expressions from sec. 2 of Appendix A  of \cite{Sangouard2011_review} truncated at $n = K$. 
We then obtain the estimated RKR in the same way as the success rate is calculated in \refsec{sec:compare_direct}: $\mathrm{RKR}_K = 1/(\overline n T+T_{swap})$ where  $T = T_0+L/c$ and $T_{swap}$ is the time  for the feed-forward. 
We also calculate the average distributed state density matrix $\rho$ as a weighted sum $\rho = \sum_{k=1}^K{P_k\rho^\prime_k}$.

In stage three, we calculate the SKF for the averaged state $\rho$ in the same way as we do for $\rho^\prime_k$ and finally obtain the SKR value as $\mathrm{SKR} = \mathrm{RKR}_K\times \mathrm{SKF}(\rho)$.

We calculate the SKR as a function $L$ for two sets of model parameters. The first set corresponds to the current experimental repeater-node realization and reads  $[0.018, 62$ ms, 0.96, 0.95, $L$, {$0.0173$ km$^{-1}$}, 123 $\mu$s, 2157 $\mu$s$]$. The second set represents the envisioned `enhanced' repeater-node with the parameters  $[0.21,630$ ms, 0.99, 0.99, $L$, $0.0173$ km$^{-1}$,123 $\mu$s, 2157 $\mu$s$]$. 
We compare the calculated SKR with the theoretical upper bound  $\mathrm{SKR}_{bound}$ determined in  \cite{Pirandola2017},  which we calculate for the attempt rate that is limited by the photon-travel-time. Specifically we use $\mathrm{SKR}_{bound} = -\log_2(1-\eta)\times c/L$.
The results of the SKR calculation are presented in Figure \ref{fig:rates} together with the RKR discussed in the \refsec{sec:compare_direct}.
The SKR predicted by the model with the current experiment parameters of set one (solid blue line in Figure \ref{fig:rates}) shows neither a cross over with $\mathrm{SKR}_{bound}$ nor advantageous scaling and yields a vanishingly small $\mathrm{SKR}$ over 50 km. Indeed, the SKR calculated for the state provided in Figure 3c of the main text is statistically consistent with zero.
For the node with the enhanced parameters of set two  (solid red line in Figure \ref{fig:rates}) the SKR delivered by the repeater is higher than the bound  $\mathrm{SKR}_{bound}$ for direct transmission for all $L > 150~$ km within the range of Figure \ref{fig:rates}.
At the crossover point of $L=150~$ km we calculate the cut-off number of memory attempts $K$ of 520. 
The model in this section does not include detector background counts, which are quantified in  \ref{sec:trapandconversion}.

\begin{figure*}[h!]
	\vspace{0mm}
	\begin{center}
        \includegraphics[width=2\columnwidth]{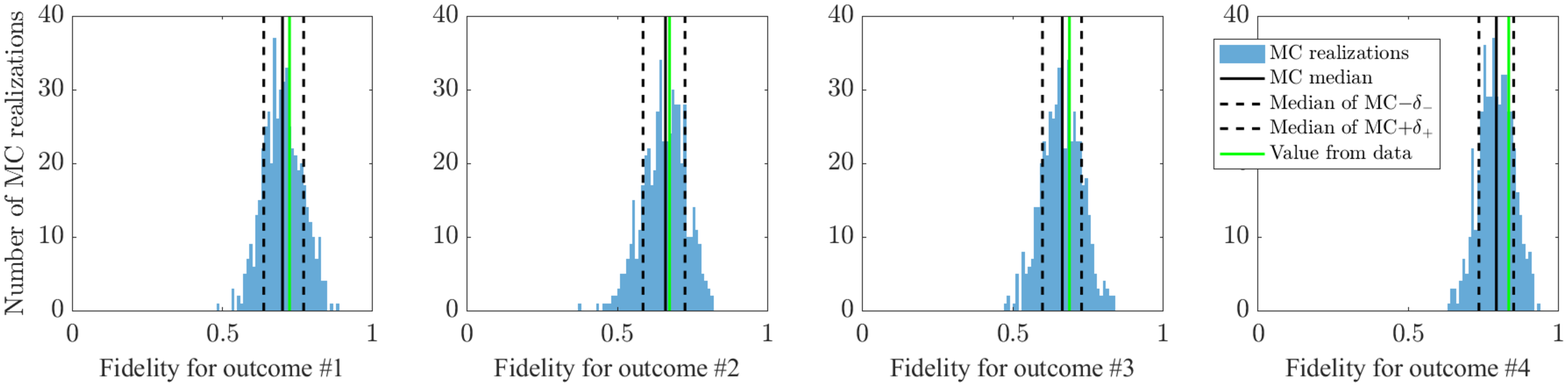}
		\vspace{0mm}
		\caption{
			\textbf{Fidelity $F_{B,i}^{Bell}$ between the measured photon-photon states $\rho^{Bell}_{B,i}$, when ion B stored a qubit in memory, and corresponding Bell states $\ket{B_i}$ when ion B stored a qubit in memory}. Four panels correspond to ion outcomes $i=1..4$. In each panel green line shows the fidelity value for the state reconstructed from the experimental data set. Bars show the distribution of fidelity values obtained theough the Monte-Carlo (MC) procedure for 500 noisy data sets. Solid black line shows the median  of that distribution. Dashed lines show the points  of the distribution  that were used to define uncertainties $\delta_-$ and $\delta_+$ as described in subsection \ref{sec:probs_and_MC}.  
			}
		\label{figsup:fid_sep2}
		\vspace{2mm}
	\end{center}
\end{figure*}

\begin{figure*}[h!]
	\vspace{0mm}
	\begin{center}
        \includegraphics[width=1\textwidth]{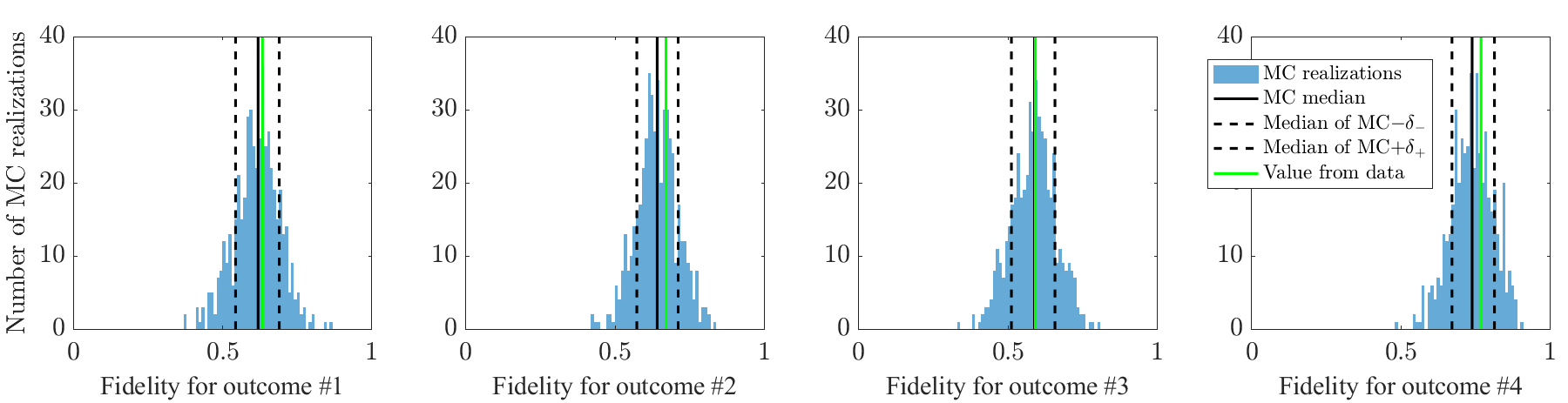}
		\vspace{0mm}
		\caption{
			\textbf{Fidelity $F_{A,i}^{Bell}$ between the measured photon-photon states $\rho^{Bell}_{A,i}$, when ion A stored a qubit in memory, and corresponding Bell states $\ket{B_i}$ }. Four panels correspond to ion outcomes $i=1..4$. The green line in each panel shows the fidelity value for the state reconstructed from the experimental data set. The blue bars of the histogram in each panel show the distribution of the  fidelity values obtained through the Monte-Carlo (MC) procedure for 500 noisy data sets. Solid black lines show the median  of that distribution. Dashed lines show the points  of the distribution  that were used to define uncertainties $\delta_-$ and $\delta_+$ as described in subsection \ref{sec:probs_and_MC}. 			}
		\label{figsup:fid_sep1}
		\vspace{2mm}
	\end{center}
\end{figure*}

\begin{figure*}[]
	\vspace{0mm}
	\begin{center}
        \includegraphics[width=1\textwidth]{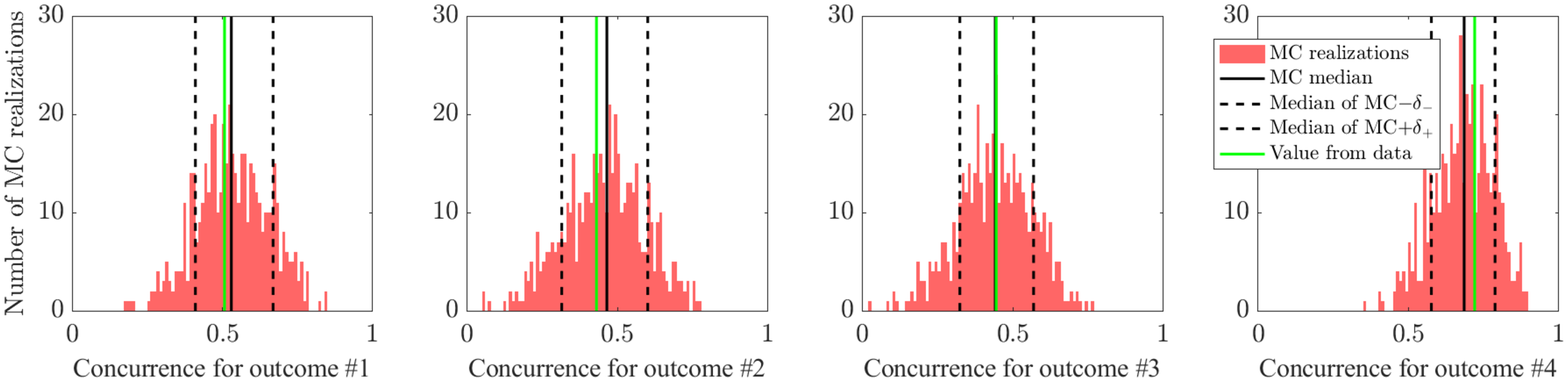}
		\vspace{-5mm}
		\caption{
			\textbf{Concurrence of  the measured photon-photon states $\rho^{B,i}$, when ion B stored a qubit in memory.} Four panels correspond to ion outcomes $i=1..4$. The green line in each panel shows the concurrence value for the state reconstructed from the experimental data set. The red bars of the histogram in each panel show the distribution of the  concurrence values obtained through the Monte-Carlo (MC) procedure for 500 noisy data sets. Solid black lines show the median  of that distribution. Dashed lines show the points  of the distribution  that were used to define uncertainties $\delta_-$ and $\delta_+$ as described in subsection \ref{sec:probs_and_MC}. 
			}
		\label{figsup:conc_sep2}
		\vspace{2mm}
	\end{center}
\end{figure*}

\begin{figure*}[]
	\vspace{0mm}
	\begin{center}
        \includegraphics[width=1\textwidth]{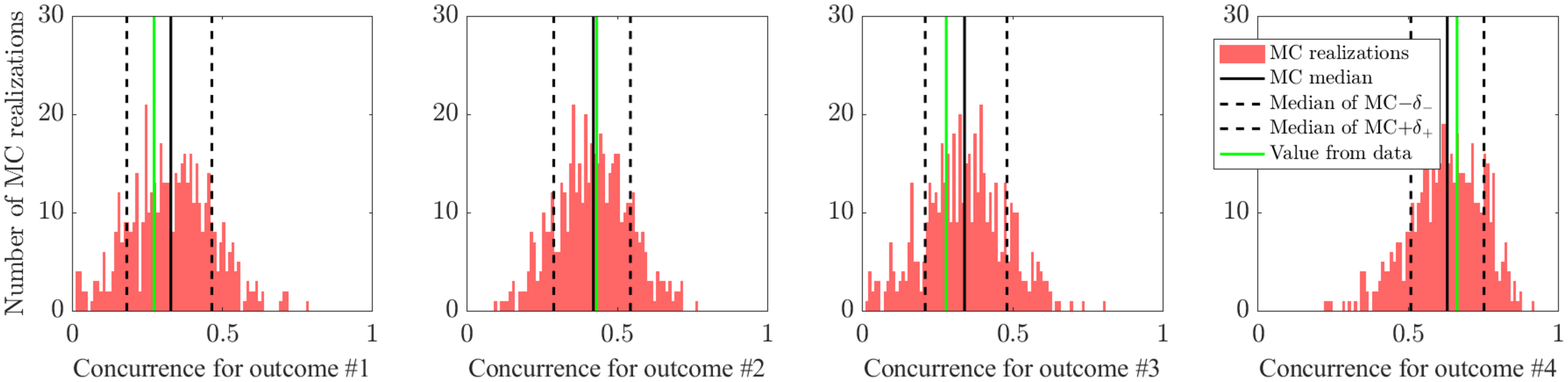}
		\vspace{-5mm}
		\caption{
			\textbf{Concurrence of  the measured photon-photon states $\rho^{A,i}$, when ion A stored a qubit in memory.}. Four panels correspond to ion outcomes $i=1..4$. The green line in each panel shows the concurrence value for the state reconstructed from the experimental data set. The red bars of the histogram in each panel show the distribution of the  concurrence values obtained through the Monte-Carlo (MC) procedure for 500 noisy data sets. Solid black lines show the median  of that distribution. Dashed lines show the points  of the distribution  that were used to define uncertainties $\delta_-$ and $\delta_+$ as described in subsection \ref{sec:probs_and_MC}. 
			}
		\label{figsup:conc_sep1}
		\vspace{2mm}
	\end{center}
\end{figure*}

\bibliography{repeater_bib.bib}